\documentclass[12pt]{article}

\usepackage{longtable}
\usepackage{amsmath}
\usepackage{amssymb}
\usepackage{algorithm}
\usepackage{latexsym}
\usepackage{epsfig}
\usepackage{epsf}
\usepackage{fullpage}
\usepackage{setspace}
\usepackage{multirow}
\usepackage{dcolumn}
\usepackage{url}

\DeclareMathOperator*{\prodkK}{\prod_{k=1}^K}

\DeclareMathOperator*{\tauk}{\tau_k^2}

\DeclareMathOperator*{\tauo}{\tau_1^2}

\DeclareMathOperator*{\tauK}{\tau_K^2}

\DeclareMathOperator*{\lamo}{\lambda_1^2}
\DeclareMathOperator*{\lamt}{\lambda_2^2}

\makeatletter     
\newcommand{\distas}[1]{\mathbin{\overset{#1}{\kern\z@\sim}}}%
\newsavebox{\mybox}\newsavebox{\mysim}
\newcommand{\distras}[1]{%
  \savebox{\mybox}{\hbox{\kern3pt$\scriptstyle#1$\kern3pt}}%
  \savebox{\mysim}{\hbox{$\sim$}}%
  \mathbin{\overset{#1}{\kern\z@\resizebox{\wd\mybox}{\ht\mysim}{$\sim$}}}%
}
\makeatother

\DeclareMathOperator*{\argmin}{arg\;min}  

\DeclareMathOperator*{\mlikelihood}{\sum_{\ell = 1}^n || \mathbf{W}^T\mathbf{x}_{\ell} - \mathbf{y}_{\ell}||_2^2}

\DeclareMathOperator*{\W}{\mathbf{W}}

\DeclareMathOperator*{\Wxl}{\mathbf{W}^T\mathbf{x}_{\ell}}
\DeclareMathOperator*{\yl}{\mathbf{y}_{\ell}}

\DeclareMathOperator*{\sig}{\sigma^2}

\DeclareMathOperator*{\vtau}{\underset{\sim}{\mathbf{\tau}^2}}

\DeclareMathOperator*{\prodink}{\prod_{i \in \pi_k}}

\DeclareMathOperator*{\sumln}{\sum_{\ell=1}^n}

\DeclareMathOperator*{\sumink}{\sum_{i \in \pi_k}}
\DeclareMathOperator*{\sumjc}{\sum_{j=1}^c}
\DeclareMathOperator*{\sumid}{\sum_{i=1}^d}

\DeclareMathOperator*{\vmuk}{\underset{\sim}{\mu_k}}

\DeclareMathOperator*{\omegai}{\omega_i^2}
\DeclareMathOperator*{\omegad}{\omega_d^2}
\DeclareMathOperator*{\omegao}{\omega_1^2}
\def \ve{\mbox{\boldmath $e$ \unboldmath}\!\!}
\def \vZ{\mbox{\boldmath $Z$ \unboldmath}\!\!}
\def \vR{\mbox{\boldmath $R$ \unboldmath}\!\!}
\def \vI{\mbox{\boldmath $I$ \unboldmath}\!\!}
\def \vW{\mbox{\boldmath $W$ \unboldmath}\!\!}
\def \vB{\mbox{\boldmath $B$ \unboldmath}\!\!}
\def \vF{\mbox{\boldmath $F$ \unboldmath}\!\!}
\def \vp{\mbox{\boldmath $p$ \unboldmath}\!\!}
\def \vU{\mbox{\boldmath $U$ \unboldmath}\!\!}
\def \vV{\mbox{\boldmath $V$ \unboldmath}\!\!}
\def \vD{\mbox{\boldmath $D$ \unboldmath}\!\!}
\def \vSigma{\mbox{\boldmath $\Sigma$ \unboldmath}\!\!}
\def \vTheta{\mbox{\boldmath $\Theta$ \unboldmath}\!\!}
\def \vLambda{\mbox{\boldmath $\Lambda$ \unboldmath}\!\!}
\def \vomega{\mbox{\boldmath $\omega$ \unboldmath}\!\!}
\def \vbeta{\mbox{\boldmath $\beta$ \unboldmath}\!\!}
\def \veta{\mbox{\boldmath $\eta$ \unboldmath}\!\!}
\def \valpha{\mbox{\boldmath $\alpha$ \unboldmath}\!\!}
\def \vgamma{\mbox{\boldmath $\gamma$ \unboldmath}\!\!}
\def \vepsilon{\mbox{\boldmath $\epsilon$ \unboldmath}\!\!}
\def \vrho{\mbox{\boldmath $\rho$ \unboldmath}\!\!}
\def \va{\mbox{\boldmath $a$ \unboldmath}\!\!}
\def \vs{\mbox{\boldmath $s$ \unboldmath}\!\!}
\def \vai{\mbox{\boldmath $a^{(i)}$ \unboldmath}\!\!}
\def \vu{\mbox{\boldmath $u$ \unboldmath}\!\!}
\def \vh{\mbox{\boldmath $h$ \unboldmath}\!\!}
\def \vb{\mbox{\boldmath $b$ \unboldmath}\!\!}
\def \vbs{\mbox{\boldmath $b(s)$ \unboldmath}\!\!}
\def \vbsi{\mbox{\boldmath $b(s_{i})$ \unboldmath}\!\!}
\def \vZ{\mbox{\boldmath $Z$ \unboldmath}\!\!}
\def \vA{\mbox{\boldmath $A$ \unboldmath}\!\!}
\def \vZss{\mbox{\boldmath $z^{*}(s)$ \unboldmath}\!\!}
\def \vZsi{\mbox{\boldmath $z(s_{i})$ \unboldmath}\!\!}
\def \vt{\mbox{\boldmath $t$ \unboldmath}\!\!}
\def \vM{\mbox{\boldmath $M$ \unboldmath}\!\!}
\def \vE{\mbox{\boldmath $E$ \unboldmath}\!\!}
\def \vTheta{\mbox{\boldmath $\Theta$ \unboldmath}\!\!}
\def \vtau{\mbox{\boldmath $\tau$ \unboldmath}\!\!}
\def \vomega{\mbox{\boldmath $\omega$ \unboldmath}\!\!}
\def \vmu{\mbox{\boldmath $\mu$ \unboldmath}\!\!}
\def \vmuZ{\mbox{\boldmath $\mu_{z}$ \unboldmath}\!\!}
\def \vpsi{\mbox{\boldmath $\psi$ \unboldmath}\!\!}
\def \vK{\mbox{\boldmath $K$ \unboldmath}\!\!}
\def \vY{\mbox{\boldmath $Y$ \unboldmath}\!\!}
\def \vL{\mbox{\boldmath $L$ \unboldmath}\!\!}
\def \vS{\mbox{\boldmath $S$ \unboldmath}\!\!}
\def \vx{\mbox{\boldmath $x$ \unboldmath}\!\!}
\def \vX{\mbox{\boldmath $X$ \unboldmath}\!\!}
\def \vWu{\mbox{\boldmath $w(u)$ \unboldmath}\!\!}

\begin{document}

\author{Keelin Greenlaw$^{1}$, Elena Szefer\,$^{2}$, Jinko Graham\,$^{2}$, Mary Lesperance,$^{1}$\\ and Farouk S. Nathoo\,$^{1,*}$\\
$^{1}$Department of Mathematics and Statistics, University of Victoria\\
$^{2}$Statistics and Actuarial Science, Simon Fraser University\\
$^{*}$nathoo@uvic.ca\\}

\title{A Bayesian Group Sparse Multi-Task Regression Model for Imaging Genetics}

\maketitle

\begin{abstract}

Motivation:
Recent advances in technology for brain imaging and high-throughput genotyping have motivated studies examining the influence of genetic variation on brain structure. Wang et al. (\textit{Bioinformatics}, 2012) have developed an approach for the analysis of imaging genomic studies using penalized multi-task regression with regularization based on a novel group $l_{2,1}$-norm penalty which encourages structured sparsity at both the gene level and SNP level. While incorporating a number of useful features, the proposed method only furnishes a point estimate of the regression coefficients; techniques for conducting statistical inference are not provided. A new Bayesian method is proposed here to overcome this limitation. 

Results:
We develop a Bayesian hierarchical modeling formulation where the posterior mode corresponds to the estimator proposed by Wang et al. (\textit{Bioinformatics}, 2012), and an approach that allows for full posterior inference including the construction of interval estimates for the regression parameters. We show that the proposed hierarchical model can be expressed as a three-level Gaussian scale mixture and this representation facilitates the use of a Gibbs sampling algorithm for posterior simulation. 
Simulation studies demonstrate that the interval estimates obtained using our approach achieve adequate coverage probabilities that outperform those obtained from the nonparametric bootstrap. Our proposed methodology is applied to the analysis of neuroimaging and genetic data collected as part of the Alzheimer's Disease Neuroimaging Initiative (ADNI), and this analysis of the ADNI cohort demonstrates clearly the value added of incorporating interval estimation beyond only point estimation when relating SNPs to brain imaging endophenotypes. Software is publicly available at \emph{https://cran.r-project.org/web/packages/bgsmtr/index.html}.

\end{abstract}

\section{Introduction}

Imaging genetics involves the use of structural or functional neuroimaging data to study subjects carrying genetic risk variants that may relate to neurological disorders such as Alzheimer's disease. In such studies the primary interest lies with examining associations between genetic variations and neuroimaging measures which represent quantitative traits. Compared to studies examining more traditional phenotypes such as case-control status, the endophenotypes derived through neuroimaging are in some cases considered closer to the underlying etiology of the disease being studied, and this may lead to easier identification of the important genetic variations. {A number of settings for statistical analysis in imaging genetics have been studied involving different combinations of gene versus genome-wide and region of interest (ROI) versus image-wide analysis, all of which have different advantages and limitations as discussed in Ge at al. (2013).

The earliest methods developed for imaging genomics data analysis are either based on significant reductions to both data types or they employ full brain-wide genome-wide scans based on a massive number of pairwise univariate analyses (e.g. Stein et al., 2010). While these approaches are convenient in terms of their implementation they ignore potential multicollinearity arising from variants within the same LD block, and they also ignore the potential relationship between the different neuroimaging endophenotypes. Ignoring these relationships precludes the borrowing of information about the genetic associations across components of the response vector. Hibar et al. (2011) use gene-based multi-variate statistics and avoid having collinearity of SNP vectors by using dimensionality reduction. Vounou et al. (2010) develop a sparse reduced-rank regression approach for studies involving high-dimensional neuroimaging phenotypes, while  Ge et al. (2012) develop a flexible multi-locus approach based on least squares kernel machines. In the latter case, the authors employ permutation testing procedures and take advantage of the spatial information inherent in brain images by using random field theory as an inferential tool (Worsley, 2002). More recently, Stingo et al. (2013) develop a Bayesian hierarchical mixture model for relating brain connectivity to genetic information for studies involving functional magnetic resonance imaging (fMRI) data. The mixture components of the proposed model correspond to the classification of the study subjects into subgroups, and the allocation of subjects to these mixture components is linked to genetic covariates with regression parameters assigned spike-and-slab priors. The proposed model is used to examine the relationship between functional brain connectivity based on fMRI data and genetic variation. 

In contrast, the focus of our work concerns the development of methodology for studies where the neuroimaging phenotypes consist of volumetric and cortical thickness measures derived from MRI which summarize the structure (as opposed to the function) of the brain over a relatively moderate number (e.g. up to 100) ROI's, and we are interested in relating brain structure to genetics. 

We develop a Bayesian approach based on a continuous shrinkage prior that encourages sparsity and induces dependence in the regression coefficients corresponding to SNPs within the same gene, and across different components of the imaging phenotypes. Our approach is related to the Bayesian group lasso (Park and Casella, 2008; Kyung et al., 2010) but it is adapted to accommodate multivariate phenotypes and it is extended to allow for grouping penalties both at the gene and SNP level. Our work is primarily motivated by the recent work of Wang et al. (2012) who propose an estimator based on group sparse regularization applied to multivariate regression where SNPs are grouped by genes or LD blocks. In what follows we will assume for specificity that the groups correspond to genes; however, this assumption is not necessary and any approach for grouping the SNPs (e.g. LD blocks) may be used. Let $\yl = (y_{\ell 1}, \dots , y_{ \ell c})^{T} \hspace{5pt}$ denote the imaging phenotype summarizing the structure of the brain over $c$ ROIs for subject $\ell$, $\ell = 1,\dots, n$. The corresponding genetic data are denoted by $\mathbf{x}_{\ell} = (x_{\ell 1}, \dots , x_{\ell d})^{T}, \hspace{5pt}\ell = 1,\dots, n$, where we have information on $d$ SNPs, and $x_{\ell j} \in \{0,1,2\}$ is the number of minor alleles for the $j^{th}$ SNP. We further assume that the set of SNPs can be partitioned into $K$ groups, for example $K$ genes, and we let  $\pi_k, k = 1,2, \dots, K$, denote the set containing the SNP indices corresponding to the $k^{th}$ group and $m_{k} = |\pi_{k}|$. We assume that $E(\mathbf{y_{\ell}}) = \mathbf{W}^{T}\mathbf{x}_\ell, \hspace{5pt} \ell = 1,\dots, n$, where $\mathbf{W}$ is a $d$ x $c$ matrix, with each row characterizing the association between a given SNP and the brain summary measures across all ROIs. The estimator proposed by Wang et al. (2012) takes the form
\begin{equation}
\label{Wang estimator} 
\hat{\mathbf{W}} = \underset{\mathbf{W}}\argmin   \mlikelihood  +\gamma_1 ||\mathbf{W}||_{G_{2,1}} +\gamma_2 ||\mathbf{W}||_{l_{2,1}}   
\end{equation}
where $\gamma_1$ and $\gamma_2$ are regularization parameters weighting a $G_{2,1}$-norm penalty \\ $||\mathbf{W}||_{G_{2,1}} = \sum_{k=1}^K  \sqrt{\sum_{i \in \pi_k}  \sum_{j=1}^c w_{ij}^2 }$ and an $\ell_{2,1}$-norm penalty $||\mathbf{W}||_{l_{2,1}} = \sum_{i=1}^{d}  \sqrt{ \sum_{j=1}^c w_{ij}^2}$ respectively. The $G_{2,1}$-norm addresses group-wise association between SNPs and encourages sparsity at the gene level. This regularization differs from group lasso (Yuan and Lin, 2006) as it penalizes regression coefficients for a group of SNPs across all imaging phenotypes jointly. As an important gene/group may contain irrelevant individual SNPs, or a less important
group may contain individually significant SNPs, the second penalty, an $\ell_{2,1}$-norm (Evgeniou and Pontil, 2007), is added to allow for additionall structured sparsity.

The estimator (\ref{Wang estimator}) provides a novel approach for assessing associations between neuroimaging phenotypes and genetic variations as it accounts for several interrelated structures within genotyping and imaging data. The incorporation of biological group structure in regression analysis with genetic data has been developed in a variety of contexts (see e.g. Stingo et al., 2011; Wen, 2014; Rockova et al., 2014; Zhu et al., 2014) . Wang et al. (2012) show that such an approach when applied to imaging genetics is able to achieve enhanced predictive performance and improved SNP selection compared with a number of alternative approaches in certain settings. Notwithstanding these advantages, a limitation of the proposed methodology is that it only furnishes a point estimate $\hat{\mathbf{W}}$ and techniques for obtaining valid standard errors or interval estimates are not provided. The primary contribution of this article is to provide an approach for doing this. 

Resampling methods such as the bootstrap are a natural starting point for this problem; however, as discussed in Kyung et al. (2010) the bootstrap estimates of the standard error for the lasso or lasso variations such as the estimator (\ref{Wang estimator}) might be unstable and not perform well.  An alternative way forward is to exploit the connection between penalized regression methods and hierarchical modeling formulations. Following the ideas of Park and Casella (2008) and Kyung et al. (2010) we develop a hierarchical Bayesian model that allows for full posterior inference. The spread of the posterior distribution then provides valid measures of posterior variability along with credible intervals for each regression parameter. Along similar lines, Bae and Mallick (2004) develop a two-level hierarchical model for gene selection that incorporates the univariate Laplace distribution as a prior that favors sparsity and employ the representation of the Laplace distribution as a Gaussian scale mixture in their model hierarchy. In our work, we use a multivariate prior based on a Gaussian scale mixture representation which is assigned independently to the set of coefficients corresponding to each gene. The prior is chosen so that the corresponding posterior mode is exactly the Wang et al. (2012) estimator. To our knowledge this specific form of multivariate shrinkage prior has not been considered previously, though the formulation is related to the general ideas developed in Kyung et al. (2010). 

The remainder of the paper proceeds as follows. In Section 2 we specify the hierarchical model and its motivation based on the estimator (\ref{Wang estimator}). The scale mixture representation is specified and a Gibbs sampling algorithm for computing the posterior distribution is presented. Section 3 presents a study of computation time and scaling, while simulation studies are presented in Section 4. Section 5 applies our methodology to a dataset obtained from the Alzheimer's Disease Neuroimaging Initiative (ADNI) database, where we relate MRI based structural brain summaries at 56 ROIs to 486 SNPs belonging to 33 genes. The final section concludes with a discussion of potential model extensions.

\section{Methods}

Let $\mathbf{W}^{(k)} = (w_{ij})_{ i \in \pi_k}$ denote the $m_{k} \times c$ submatrix of $\mathbf{W}$ containing the rows corresponding to the $k^{th}$ gene, $k=1, \dots, K$. The hierarchical model corresponding to the estimator (\ref{Wang estimator}) takes the form
\begin{equation}
\label{model - level 1}
\yl |\mathbf{W},\sigma^2  \distas{ind} MVN_c (\vW^{T} \vx_\ell \: , \: \sigma^2I_c) \hspace{8pt} \ell=1, \dots, n, 
\end{equation}
with the coefficients corresponding to different genes assumed conditionally independent
\begin{equation}
\label{model -level 2} 
\mathbf{W}^{(k)}| \lambda_{1}^{2}, \lambda_{2}^{2}, \sigma^2  \distas{ind} p(\mathbf{W}^{(k)}|  \lambda_{1}^{2}, \lambda_{2}^{2}, \sigma^2) \hspace{8pt} k=1,\dots,K, 
\end{equation}
and with the prior distribution for each $\mathbf{W}^{(k)}$ having a density function given by
\begin{equation}
\label{PML}
\begin{split}
p(\mathbf{W}^{(k)} |  \lambda_{1}^{2}, \lambda_{2}^{2}, \sigma^2) \propto \exp \left\lbrace - \frac{\lambda_{1}}{\sigma} \sqrt{ \sum_{i \in \pi_k} \sum_{j=1}^c w_{ij}^2 } \right\rbrace \\ \times \prod_{i \in \pi_k} \exp \left\lbrace -\frac{\lambda_{2}}{\sigma} \sqrt{\sum_{j=1}^c w_{ij}^2 } \right\rbrace.
\end{split}
\end{equation}
The shrinkage prior (\ref{PML}) is not a multivariate Laplace distribution; however, each term of the product on the right-hand side of (\ref{PML}) is the kernel of a form of the multivariate Laplace distribution discussed in Kotz et al. (2001), and so we refer to this prior as the \emph{product multivariate Laplace distribution}. The prior is specified conditional on $\sigma$ and the dependence of the prior density on $\sigma$ follows the parameterization of the univariate Laplace distribution considered in Park and Casella (2008) who show that this parameterization guarantees a unimodal posterior for the Bayesian lasso. By construction, the posterior mode, conditional on $ \lambda_{1}^{2}, \lambda_{2}^{2}, \sigma^2$,  corresponding to the model hierarchy (2) - (4) is exactly the estimator (1) proposed by Wang et al. (2012) with $\gamma_{1} = 2\sigma \lambda_{1}$ and $\gamma_{2} = 2  \sigma \lambda_{2}$. This equivalence between the posterior mode and the estimator of Wang et al. (2012) is the motivation for our model; however, we note that generalizations that allow for a more flexible covariance structure in (\ref{model - level 1}) could also be considered. For the current model each component of $\yl$ is scaled to have unit variance across subjects, making the assumption of a single variance component $\sigma^{2}$ tenable. We also note that while (\ref{model - level 1}) assumes conditional independence across imaging phenotypes, the prior distribution (\ref{PML}) induces dependence in the regression coefficients across the imaging phenotypes for coefficients corresponding to the same gene (group).\\

\noindent PROPOSITION 1. (Prior Propriety) \emph{The prior for $\vW$ based on (\ref{model -level 2}) and (\ref{PML}) is proper.}\\ 

{\bf Proof:} For each $k \in \{1, \dots, K\}$ we define $I_{k} $ as
\begin{equation*}
\begin{split}
&I_{k} = \int \exp \left\lbrace - \frac{\lambda_{1}}{\sigma} \sqrt{ \sum_{i \in \pi_k} \sum_{j=1}^c w_{ij}^2 } \right\rbrace\\
&\times \prod_{i \in \pi_k} \exp \left\lbrace -\frac{\lambda_{2}}{\sigma} \sqrt{\sum_{j=1}^c w_{ij}^2 } \right\rbrace d \mathbf{W}^{(k)}.
\end{split}
\end{equation*}
It is sufficient to show that $\prod_{k=1}^{K} I_{k}$ is finite. We note that 
\begin{equation}
\label{upper_bound}
I_{k} \le \int \exp \left\lbrace - \frac{\lambda_{1}}{\sigma} \sqrt{ \sum_{i \in \pi_k} \sum_{j=1}^c w_{ij}^2 } \right\rbrace  d \mathbf{W}^{(k)}
\end{equation}
since $\exp(-x) \le 1$ for $x \ge 0$. The integrand on the right-hand-side of (\ref{upper_bound}) is proportional to the probability density function of a particular form of the multivariate Laplace distribution discussed in Kotz et al. (2001). Given this form, the integral can be evaluated as
\begin{equation*}
\begin{split}
&\int \exp \left\lbrace - \frac{\lambda_{1}}{\sigma} \sqrt{ \sum_{i \in \pi_k} \sum_{j=1}^c w_{ij}^2 } \right\rbrace  d \mathbf{W}^{(k)} = \pi^{(m_{k}c-1)/2}\\
&\times \Gamma((m_{k}c+1)/2)2^{m_{k}c}(\lambda_{1}^{2}/\sigma^{2})^{-m_{k}c/2} < \infty,
\end{split}
\end{equation*}
so that $I_{k} < \infty$ and therefore $\prod_{k=1}^{K} I_{k} < \infty$ as required.\\

If the hyper-parameters $\sigma^{2}$, $\lambda_{1}$, and $\lambda_{2}$ are fixed or assigned proper priors then Proposition 1 is sufficient to ensure that the posterior distribution is proper. The following proposition provides a stochastic representation of the prior based on a Gaussian scale mixture. This representation is important as it facilitates computation of the posterior distribution using a simple Gibbs sampling algorithm.\\

\noindent PROPOSITION 2. (Scale mixture representation) \emph{For each\\ $i \in \{1,\dots,d\}$ let $k(i) \in \{1,\dots,K\}$ denote the gene associated with the $i^{th}$ SNP. The prior (\ref{PML}) can be obtained through the following scale mixture representation:
\begin{equation}
\label{scale-mix1}
w_{ij}\; |\; \sigma^2, \; \vtau^{2}, \; \vomega^{2} \; \distas{ind} \; N \left( 0, \;\sigma^2 (  \frac{1}{\tau_{k(i)}^2}\; +\; \frac{1}{\omega_i^2} )^{-1}\: \right), 
\end{equation}
with continuous scale mixing variables $\vtau^{2} = (\tau_{1}^{2},\dots,\tau_{K}^{2})'$ and $\vomega^{2} = (\omega_{1}^{2},\dots,\omega_{d}^{2})'$ distributed according to the density
\begin{equation}
\label{scale-mix2}
\begin{split}
&p(\vtau^{2},\vomega^{2}|\lambda_{1}^{2},\lambda_{2}^{2})\\
&\propto \prodkK  \left(\frac{\lambda_1^2}{2}\right)^{\left(\frac{m_kc + 1}{2}\right)}(\tau_k^2)^{\left(\frac{m_kc + 1}{2}\right) -1 } \exp \left\lbrace -\left(\frac{\lambda_1^2}{2}\right) \tau_k^2 \right\rbrace \\
&\times  \prod_{i \in \pi_{k}}   \left(\frac{\lambda_2^2}{2}\right)^{\left(\frac{c + 1}{2}\right)}(\omega_i^{2})^{\left(\frac{c + 1}{2}\right) -1 } \exp \left\lbrace -\left(\frac{\lambda_2^2}{2}\right) \omega_{i}^{2} \right\rbrace\\
&\times (\tau_{k}^{2}+ \omega_{i}^{2})^{-\frac{c}{2}}.  
\end{split}
\end{equation}}

{\bf Proof:} From Kyung et al. (2010, Appendix 2) we have the following:
\begin{equation}
\label{eq:Wprior1}
\begin{split} 
&\exp \left\lbrace  -\frac{\lambda_1}{\sigma}  \:||\boldsymbol{W}^{(k)}||_2 \right\rbrace 
\propto  \int_0^{\infty} \left( \frac{1}{2\pi\sigma^2\tau_{k}^2} \right)^{\frac{m_kc}{2}} \\
&\times \exp\left\lbrace -\frac{||\boldsymbol{W}^{(k)}||_2^2}{2\sigma^2\tau_{k}^2} \right\rbrace \; 
\frac{ (\frac{\lambda_1^2}{2})^{\left(\frac{m_kc + 1}{2}\right)}}{\Gamma \left(\frac{m_kc + 1}{2}\right) } (\tau_{k}^2)^{\left(\frac{m_kc + 1}{2}\right) -1} \;\\ 
&\times \exp \left\lbrace - \left(\frac{\lambda_1^2}{2}\right)  \tau_{k}^2 \right\rbrace d \tau_{k}^2,
\end{split}
\end{equation}
and
\begin{equation}
\begin{split}
\label{eq:Wprior2}
&\exp \left\lbrace - \frac{\lambda_2}{\sigma} ||\boldsymbol{w}^i||_2 \right\rbrace  
\propto \; \int_0^\infty \left( \frac{1}{2\pi\sigma^2\omega_i^2} \right)^{\frac{c}{2}} 
\exp \left\lbrace -\frac{||\boldsymbol{w}^i||_2^2}{2\sigma^2\omega_i^2} \right\rbrace\\
& \times \frac{ (\frac{\lambda_2^2}{2})^{\left(\frac{c + 1}{2}\right)}}{\Gamma \left(\frac{c + 1}{2}\right)} (\omega_i^2)^{\left(\frac{c + 1}{2}\right) -1} \; 
 \exp \left\lbrace -  \left(\frac{\lambda_2^2}{2}\right) \omega_i^2 \right\rbrace  d\omega_i^2, 
\end{split}
\end{equation}
where $\boldsymbol{w}^i$ denotes the $i^{th}$ row of $\vW$. Beginning with (\ref{PML}) we substitute (\ref{eq:Wprior1}) and (\ref{eq:Wprior2}), apply some algebra, and simplify to obtain $p(\mathbf{W}^{(k)} |  \lambda_{1}^{2}, \lambda_{2}^{2}, \sigma^2) $
$$
\propto \int_{0}^{\infty} \cdots \int_{0}^{\infty}   \prod_{i \in \pi_k} \left[ \left(\sig \left(\frac{1}{\tau_{k}^2}\:+\:\frac{1}{\omega_i^2}\right)^{-1}\right)^{-\frac{c}{2}} \right] 
$$
$$
\times \exp \left\lbrace - \sumink \left( \frac{  \sumjc w_{ij}^2}{2\sig \left(\frac{1}{\tau_{k}^2}\:+\:\frac{1}{\omega_i^2}\right)^{-1}} \right) \right\rbrace \exp \left\lbrace -\frac{\lambda_{1}^{2}}{2} \tau_{k}^{2} \right\rbrace 
$$
$$
\times  
\left[  \prod_{i \in \pi_k}  \left(\sig \left(\frac{1}{\tau_{k}^2}\:+\:\frac{1}{\omega_i^2}\right)^{-1}\right)^{\frac{c}{2}} \right]
\times (\frac{\lambda_1^2}{2})^{\left(\frac{m_kc + 1}{2}\right)} (\tau_{k}^{2})^{-\frac{1}{2}} 
$$
$$
\times  \left[\prod_{i \in \pi_{k}} (\frac{\lambda_2^2}{2})^{\left(\frac{c + 1}{2}\right)}  (\omega_{i}^{2})^{-\frac{1}{2}} 
\exp \left\lbrace -\frac{\lambda_{2}^{2}}{2} \omega_{i}^{2} \right\rbrace d \omega_{i}^{2} \right]d \tau_{k}^{2}
$$
From (\ref{model -level 2}), we are able to take the product of the expression above over $k \in \{1,\dots,K\}$, and after simplification we obtain $p(\mathbf{W} |  \lambda_{1}^{2}, \lambda_{2}^{2}, \sigma^2)$
\begin{align}
\label{final-prop1}
\begin{split}
&\propto \int_{0}^{\infty} \cdots \int_{0}^{\infty} \prod_{k=1}^{K} \prod_{i \in \pi_{k}} N(w_{ij};0,\sig \left(\frac{1}{\tau_{k}^2}\:+\:\frac{1}{\omega_i^2}\right)^{-1})\\
&\times \prod_{k=1}^{K}(\frac{\lambda_1^2}{2})^{\left(\frac{m_kc + 1}{2}\right)}(\tau_{k}^{2})^{\frac{m_{k}c+1}{2}-1} 
\exp \left\lbrace -\frac{\lambda_{1}^{2}}{2} \tau_{k}^{2} \right\rbrace\\
& \times \left[\prod_{i \in \pi_{k}} (\frac{\lambda_2^2}{2})^{\left(\frac{c + 1}{2}\right)}  (\omega_{i}^{2})^{\frac{c+1}{2}-1} 
\exp \left\lbrace -\frac{\lambda_{2}^{2}}{2} \omega_{i}^{2} \right\rbrace \right]\\ 
&\times \left[\prod_{i \in \pi_{k}}(\tau_{k}^{2} + \omega_{i}^{2})^{-\frac{c}{2}}d \omega_{i}^{2} \right]d \tau_{k}^{2},
\end{split}
\end{align}
where $N(x;\mu,\sigma^{2})$ denotes the density of a normal distribution with mean $\mu$, variance $\sigma^{2}$ evaluated at $x$. The first line of the integrand in (\ref{final-prop1}) corresponds to (\ref{scale-mix1}), while the remaining lines of (\ref{final-prop1}) correspond to (\ref{scale-mix2}), and the integration is over the scale mixing variables $\vtau^{2}$ and $\vomega^{2}$. It follows that (\ref{model -level 2})-(\ref{PML}) can be represented through the Gaussian scale mixture (\ref{scale-mix1})-(\ref{scale-mix2}).\\

This hierarchical representation of the shrinkage prior (\ref{scale-mix2}) introduces gene specific latent variables $\tau_{1}^{2},\dots,\tau_{K}^{2}$ as well as SNP specific latent variables $\omega_{1}^{2},\dots,\omega_{d}^{2}$ that modulate the conditional variance of each regression coefficient in (\ref{scale-mix1}). Unlike other formulations for Bayesian lassos the scale mixing variables are not assumed independent. The dependence in the joint distribution arises from the term $(\tau_{k}^{2}+ \omega_{i}^{2})^{-\frac{c}{2}}$ in (\ref{scale-mix2}) and this is required to ensure that the resulting marginal distribution for $\vW$ has the required form (\ref{PML}). The parameter $\sigma^{2}$ is assigned a proper inverse-Gamma prior
\begin{equation}
\label{sigma-prior}
\sigma^2 \; \sim \; Inv-Gamma (a_\sigma, b_\sigma ),
\end{equation}
and the hierarchical model (\ref{model - level 1}), (\ref{scale-mix1}), (\ref{scale-mix2}), and (\ref{sigma-prior}) has a conjugacy structure that facilitates posterior simulation using a Gibbs sampling algorithm. As the normalizing constant associated with (\ref{scale-mix2}) is not known and may not exist, we work with the unnormalized form which yields proper full conditional distributions having standard form. Our focus of inference does not lie with the scale mixing variables themselves, rather, the use of the scale mixture representation is a computational device that leads to a fairly straightforward Gibbs sampling algorithm which enables us to draw from the marginal posterior of $\vW$. By Proposition 1 and the fact that (\ref{sigma-prior}) is proper we are assured that this posterior distribution is always proper. The Gibbs sampler is presented in Algorithm 1 while the corresponding derivations are presented in the supplementary material. Starting values for the algorithm can be obtained in part by first computing the estimator (\ref{Wang estimator}) and using these to initialize the MCMC sampler.

\vspace{-1em}
\begin{algorithm}[H]
\caption{Gibbs Sampling Algorithm}
\begin{enumerate}
\footnotesize
\item Set tuning parameters $\lambda_{1}^{2}$ and $\lambda_{2}^{2}$.
\item Initialize $\vW$, $\vtau^{2}$, $\vomega^{2}$ and repeat steps (3) - (6) below to obtain the desired Monte Carlo sample size after burn-in. 
\item Update $\sigma^{2} \sim \text{Inv-Gamma}(a_{\sigma}^{*},b_{\sigma}^{*})$,
$
a_{\sigma}^{*} = \frac{c}{2}(n+d)+a_{\sigma}
$
\begin{equation*}
\begin{split}
&b_{\sigma}^{*} = \frac{1}{2}\sum_{l=1}^{n}||\yl - \Wxl||_{2}^{2} \\
&+ \frac{1}{2}\sum_{i=1}^{d}( \frac{1}{\tau_{k(i)}^2}\; +\; \frac{1}{\omega_i^2} )\sum_{j=1}^{c}w_{ij}^{2} + b_{\sigma}.
\end{split}
\end{equation*}
\item For $k = 1, \dots,K$ update $\tau_{k}^{2}$, through\\ $1/\tau_{k}^{2} \sim \text{Inverse-Gaussian}(\sqrt{\frac{\lambda_{1}^{2}\sigma^{2}}{||\mathbf{W}^{(k)}||_{F}^{2}}}, \lambda_{1}^{2})$.
\item For $i = 1, \dots,d$ update $\omega_{i}^{2}$, through\\ $1/\omega_{i}^{2} \sim \text{Inverse-Gaussian}(\sqrt{\frac{\lambda_{2}^{2}\sigma^{2}}{\sum_{j=1}^{c}w_{ij}^{2}}},\lambda_{2}^{2})$.
\item For $k=1,\dots,K$ update $\mathbf{W}^{(k)}$, based  on\\ $\text{vec}(\mathbf{W}^ {(k)'} ) \sim MVN_{m_{k}c}(\vmu_{k},\vSigma_{k})$ where
\begin{equation*}
\begin{split}
&\vmu_{k} = -\vA_{k}^{-1}\sum_{l=1}^{n}(\vx_{\ell}^{(k)} \otimes \vI_{c})  (\vx_{\ell}^{(-k)'} \otimes \vI_{c})  \text{vec}(\mathbf{W}^ {(-k)'}) \\
&+ \vA_{k}^{-1}\sum_{l=1}^{n}  (\vx_{\ell}^{(k)} \otimes \vI_{c})\yl,\,\,\,\vSigma_{k} = \sigma^{2} \vA_{k}^{-1}, \, \vA_{k} =  
\end{split}
\end{equation*}

\begin{equation*}
\begin{split}
& \sum_{l=1}^{n}(\vx_{\ell}^{(k)} \otimes \vI_{c})   (\vx_{\ell}^{(k)'} \otimes \vI_{c})+ \text{Diag}\{\frac{1}{\tau_{k}^{2}}  + \frac{1}{\omega_{i}^{2}} \}_{ i \in \pi_{k}} \otimes  \vI_{c}    
\end{split}
\end{equation*}
and where $\mathbf{W}^ {(-k)} = (w_{ij})_{i \not\in \pi_{k},j}$, $\vx_{\ell}^{(k)} = (x_{\ell j})_{j \in \pi_{k}}$, \\and $\vx_{\ell}^{(-k)} = (x_{\ell j})_{j \not\in \pi_{k}}$.
\end{enumerate}
\end{algorithm}
\normalsize

The tuning parameters $\gamma_{1}$, $\gamma_{2}$  in (\ref{Wang estimator}) and $\lambda_{1}^{2}$, $\lambda_{2}^{2}$ in the hierarchical model (\ref{model - level 1}), (\ref{scale-mix1}), (\ref{scale-mix2}), and (\ref{sigma-prior}) control the strength of the regularization terms and thus the structure of the penalty that governs the bias-variance tradeoff associated with the estimator of $\mathbf{W}$. Wang et al. (2012) suggest the use of five-fold cross-validation (CV) over a discrete two-dimensional grid $\{10^{-5},10^{-4},\dots,10^4,10^5\}^2$ of possible values. A problem with the use of CV when MCMC runs are required to fit the model is that an extremely large number of parallel runs are needed to cover all points on the grid for each possible split of the data. To avoid some of this computational burden we approximate leave-one-subject-out CV using the WAIC (Watanabe, 2010; Gelman et al., 2014)
$$
WAIC = -2\sum_{l=1}^{n}\log E_{\W,\sigma^{2}}[p(\yl | \W,\sigma^{2}) | \mathbf{y}_{1}, \dots, \mathbf{y}_{n}] 
$$
$$
+ 2\sum_{l=1}^{n} VAR_{\W,\sigma^{2}}[\log p(\yl | \W,\sigma^{2}) | \mathbf{y}_{1}, \dots, \mathbf{y}_{n}]
$$
where $p(\yl | \W,\sigma^{2})$ is the probability density function associated with (\ref{model - level 1}) and the required posterior means and variances are approximated based on the output of the MCMC sampler at each point of the grid. These samplers are run in parallel using a high performance computing cluster. The values of $\lambda_{1}^{2}$ and $\lambda_{2}^{2}$ are then chosen as those values that minimize the WAIC across the grid and no data-splitting is required. We note that alternative approaches based on either empirical Bayes (EB) or hierarchical Bayes (HB) could also be used to choose the tuning parameters; however, for the model under consideration we have found  (Nathoo et al., 2016) that using both EB and HB to select the tuning parameters can lead to severe over-shrinkage of the posterior mean of the regression coefficients when $d>n$ or when the genetic effects are weak.

\section{Computation Time and Scaling}

In this section we report on computation times and scaling as the number of subjects $n$, the dimension of the phenotype $c$, and the number of SNPs $d$ changes. Three experiments are performed with each examining how the computation time scales with one of the three input dimensions. The computation times reported here are based on a total of 10,000 MCMC iterations (5,000 iterations was a sufficient burn-in in all cases considered) with each run employing 49 cores (each 2.66-GHz Xeon x5650) on a computing cluster with 20GB of RAM requested for each job. To be clear on the parallel aspect of the computing, each core is simply used to run the Gibbs sampler with a different value of $(\lambda_{1}^{2},\lambda_{2}^{2})$ and the value minimizing the WAIC is used for inference in each case. The computational algorithm itself runs on a single core. When multiple cores are not available, our R package 'bgsmtr' provides alternative approaches for choosing the tuning parameters with computations using only a single core.

We choose baseline values of $c=12$, $d=500$, $n=600$, and in each of the three experiments the data are simulated from the model with one dimension varying while the other two are fixed at the baseline values. The results from the three experiments are displayed in Figure 1 and Figure 2. In each case the computation time scales approximately linearly with the given input. For a fully Bayesian approach with implementation based on MCMC, the computation time is not extensive even for the most extreme values ($d=5,000$, $c=100$, $n=10,000$) and larger values can be considered if more memory is available, or alternatively, thinning can be applied to the MCMC chains to reduce the memory requirements.
%
%
\begin{figure*}[t]
\centering
\begin{tabular}{cc}
\includegraphics[scale=0.45]{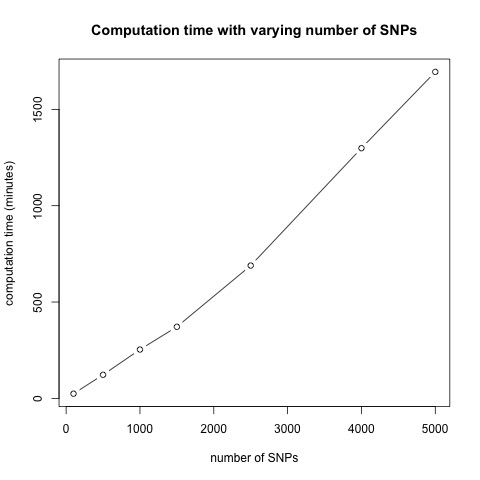} &
\includegraphics[scale=0.45]{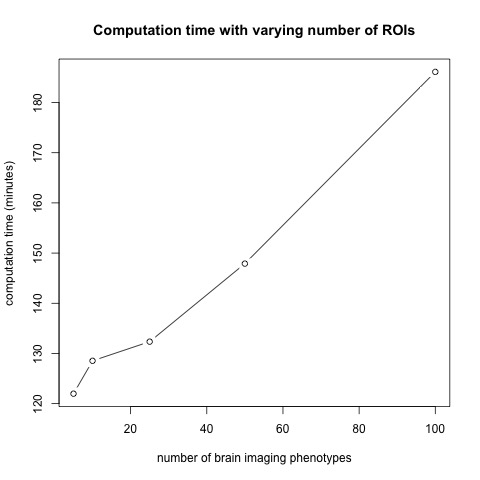} 
\end{tabular}
\caption{Computation time in minutes as a function of the number of SNPs $d$ ($c=12$, $n=600$) and the number of phenotypes $c$ ($d=500$, $n=600$).}
\end{figure*}
\begin{figure*}[!t]
\centering
\includegraphics[scale=0.45]{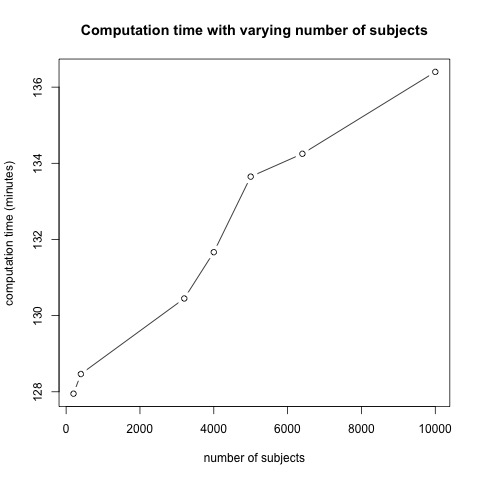}
\caption{Computation time as a function of the number of subjects $n$ ($c=12$, $d=500$).}
\end{figure*}

\section{Simulation Studies}

We conduct four simulation studies in which our proposed methodology is evaluated with the primary objective of evaluating the coverage probabilities of the 95\% equal-tail credible intervals for the regression coefficients $\vW.$ We focus on evaluating coverage probabilities as the ability to quantify uncertainty through interval estimation is the primary value-added of our methodology over and above the estimator proposed by Wang et al. (2012). We also compare our approach to a more standard approach, the nonparametric bootstrap applied to the estimator (\ref{Wang estimator}).

The application of the nonparametric bootstrap involves resampling the data with replacement and recomputing the estimator (\ref{Wang estimator}) for each bootstrap sample. The bootstrap distribution of the resulting estimators over a large number $B=1000$ bootstrap samples is then used to construct approximate 95\% confidence intervals. In this case the bootstrap resampling is done at the level of subjects. The tuning parameters $\gamma_{1}$ and $\gamma_{2}$ are recomputed for each simulated dataset in the simulation study but they are fixed across all bootstrap replicates corresponding to a single simulated dataset. The selection for these tuning parameters is based on five-fold CV. 

The simulation studies are based on genetic data obtained from the ADNI database. The data comprise information on $d=486$ SNPs belonging to $K=33$ genes obtained from a total $n=632$ subjects (179 cognitively normal (CN), 144 AlzheimerÕs disease (AD), 309 late mild cognitive impairment (LMCI) stage). The genes for which we have information along with the number of SNPs included for each gene are depicted in Figure 1 of the supplementary material. 

We include all $486$ SNPs and simulate imaging data from $c=12$ ROIs, with Study I having $n=632$ subjects, and Study II having $n=250$ ($83$ CN, $83$ AD, $84$ LMCI) subjects. Study II differs from Study I in that we move to a high-dimensional setting by reducing the value of $n$ so that $n<d$. In each case we set the true values as $\lambda_{1}^{2}=\lambda_{2}^{2}=\sigma^{2}=2$, and set the true values for $\vW$ by first
simulating 
$
\tau_k^2 \; | \; \lambda_{1}^{2} \; \distas{ind} \; Gamma \left( \frac{m_k\*c+1}{2} \;, \; \frac{\lambda_{1}^{2}}{2} \right),\,\,\, k=1,\dots,K, 
$ and $
\omega_i^2 \; | \; \lambda_{2}^{2} \; \distas{ind} \; Gamma \left( \frac{c+1}{2} \;, \; \frac{\lambda_{2}^{2}}{2} \right), \,\,\, i=1,\dots,d, 
$ and then simulating the regression coefficients from (\ref{scale-mix1}), and finally, the true values for $\vW$ are obtained by setting the entries of all but $50$ rows of $\vW$ to zero. This adds additional sparsity to the SNP effects and makes the simulation setup more realistic. We note that the simulation of $\vtau^{2}$ and $\vomega^{2}$ from Gamma distributions is not based on our assumed model and the additional sparsity added after simulation from (\ref{scale-mix1}) does not correspond to the prior from our model, so that we are not assuming that the model is correctly specified. The non-zero rows correspond to $5$ genes containing exactly $14$, $10$, $6$, $4$, and $1$ SNP(s) respectively (for a total of $35$ SNPs), along with an additional $15$ rows corresponding to additional SNPs. The imaging data are simulated from (\ref{model - level 1}) and we note that the model assumption (\ref{model - level 1}) is common to both of the approaches being compared, so neither has an advantage.

To further investigate the robustness of our approach relative to the bootstrap in settings where the model assumptions do not match the model from which the data have been generated we conduct two additional simulation studies, labelled Study III and Study IV, which have the same settings as Study I and Study II, respectively, with the exception that the regression errors are drawn from a heavy-tailed multivariate $t_{4}$ distribution.

For each of $100$ simulation replicates we compute the bootstrap 95\% confidence interval based on the estimator (\ref{Wang estimator}) and the posterior distribution from our Bayesian model using the Gibbs sampling algorithm. In total each simulation study involves $d\times c = 5,832$ regression parameters and we use the 100 simulation replicates to estimate the coverage probability of the 95\% equal-tail confidence/credible intervals for each parameter. The results are presented in Table 1.
 
\begin{table}
\footnotesize
\begin{center}
\caption{Simulation studies - interval estimation. The coverage probability of each approximate 95\% credible/confidence interval is estimated based on 100 simulation replicates and then averaged (MCP) overall and also separately over the parameters that correspond to active SNPs.} 
\begin{tabular}{lcc}
\hline
\multicolumn{3}{c}{Study I}\\
Method & MCP (overall)  & MCP ($w_{ij} \ne 0$)    \\
\hline
Bayesian Model& 0.95&0.83\\
Nonparametric Bootstrap&0.85  &0.45\\
\multicolumn{3}{c}{Study II}\\
Method & MCP (overall)  & MCP ($w_{ij} \ne 0$)   \\
\hline
Bayesian Model& 0.94&0.72\\
Nonparametric Bootstrap&0.85  &0.42\\
\multicolumn{3}{c}{Study III}\\
Method & MCP (overall)  & MCP ($w_{ij} \ne 0$)    \\
\hline
Bayesian Model& 0.97 &0.77\\
Nonparametric Bootstrap&0.86  &0.49\\
\multicolumn{3}{c}{Study IV}\\
Method & MCP (overall)  & MCP ($w_{ij} \ne 0$)   \\
\hline
Bayesian Model& 0.95& 0.73\\
Nonparametric Bootstrap&0.84  &0.41\\
\hline

\end{tabular}
\end{center}
\normalsize
\end{table}

In Study I we find that the mean (over all $5,832$ parameters) coverage probability is 95\% for intervals constructed based on our approach, while that for the nonparametric bootstrap applied to the estimator of Wang et al. (2012) is 85\%, below the nominal level. Considering only those $600$ parameters with non-zero effects the mean coverage probability for our approach drops to 83\%, while that for the nonparametric bootstrap drops to an unreasonable 45\%. In Study II ($n<d$) we find that the mean (over all $5,832$ parameters) coverage probability is 94\% for our approach while that obtained for intervals constructed using the nonparametric bootstrap is 85\%. Considering only those parameters with non-zero true values the mean coverage probabilities associated with both approaches drops as in Study I, to 72\% for our approach and to 42\% for the nonparametric bootstrap. The results for Studies III and IV generally indicate the same patterns as those seen in Studies I and II, demonstrating that our comparisons exhibit some robustness to model misspecification.

We find that the Bayesian approach is clearly outperforming the estimator of Wang et al. (2012) combined with the nonparametric bootstrap in all cases. In all four studies the mean coverage probability drops when considering only active SNPs, but in this case the values obtained from the nonparametric bootstrap are unreasonably low while those obtained from our approach are still somewhat reasonable, in particular since these coverage probabilities pertain to active SNPs, and therefore, under-coverage will not lead to a false rejection of the null hypothesis. 

\section{Application to ADNI Data}
We illustrate our methodology by applying it to a dataset obtained from the Alzheimer's Disease Neuroimaging Initiative (ADNI-1) database. This dataset includes both genetic and structural MRI data and is similar to a dataset analyzed by Wang et al. (2012); however, we use a larger number of regions of interest in our analysis leading to 56 imaging phenotypes rather than the 12 imaging phenotypes analyzed by Wang et al. (2012). The imaging phenotypes used in our analysis are listed in Table 2.
\begin{table*}[t]
\begin{center}
\caption{
Imaging phenotypes defined as volumetric or cortical thickness measures of
$28 \times 2 = 56$ regions of interest (ROIs) from automated Freesurfer parcellations. Each of the phenotypes in the table corresponds to two phenotypes in the data: one for the left hemisphere and the other for the right hemisphere.} 
\tiny
\begin{tabular}{lll}
\hline
ID & Measurement & Region of interest\\
AmygVol & Volume&Amygdala\\
CerebCtx & Volume&Cerebral cortex\\
CerebWM & Volume&Cerebral white matter\\
HippVol & Volume&Hippocampus\\
InfLatVent &Volume&Inferior lateral ventricle\\ 
LatVent &Volume&Lateral ventricle\\
EntCtx &Thickness&Entorhinal cortex\\
Fusiform &Thickness&Fusiform gyrus\\
InfParietal &Thickness&Inferior parietal gyrus\\
InfTemporal &Thickness&Inferior temporal gyrus\\ 
MidTemporal &Thickness&Middle temporal gyrus\\ 
Parahipp &Thickness&Parahippocampal gyrus\\
PostCing &Thickness&Posterior cingulate\\
Postcentral &Thickness&Postcentral gyrus\\ 
Precentral &Thickness&Precentral gyurs\\
Precuneus &Thickness&Precuneus\\
SupFrontal &Thickness&Superior frontal gyrus\\
SupParietal &Thickness&Superior parietal gyrus\\
SupTemporal &Thickness&Superior temporal gyrus\\
Supramarg &Thickness&Supramarginal gyrus\\
TemporalPole &Thickness&Temporal pole\\

MeanCing&Mean thickness&Caudal anterior cingulate, isthmus cingulate,\\ 
&&posterior cingulate,rostral anterior cingulate\\

MeanFront&Mean thickness& Caudal midfrontal, rostral midfrontal, superior frontal,\\ 
&& lateral orbitofrontal, and medial orbitofrontal gyri, frontal pole\\

MeanLatTemp&Mean thickness&Inferior temporal, middle temporal, and superior temporal gyri\\

MeanMedTemp&Mean thickness&Fusiform, parahippocampal, and lingual gyri,\\
&& temporal pole and transverse temporal pole\\

MeanPar&Mean thickness&Inferior and superior parietal gyri, supramarginal gyrus,\\ 
&&and precuneus\\
MeanSensMotor&Mean thickness&Precentral and postcentral gyri\\ 
MeanTemp&Mean thickness&Inferior temporal, middle temporal, superior temporal,\\
&& fusiform, parahippocampal, lingual gyri, temporal pole,\\ 
&&transverse temporal pole\\
\hline
\end{tabular}
\end{center}
\label{default}
\end{table*}%

Registered ADNI investigators may obtain the preprocessed data used in this analysis by contacting the corresponding author. These data can be used in conjunction with our R package 'bgsmtr' implementing our methodology to reproduce the results presented here. 

The data are available for $n=632$ subjects (179 CN, 144 AD, 309 LMCI), and among all possible SNPs we include only those SNPs belonging to the top 40 Alzheimer's Disease (AD) candidate genes listed on the AlzGene database as of June 10, 2010. The data presented here are queried from the most recent genome build as of December 2014, from the ADNI-1 data. 

After quality control and imputation steps, the genetic data used for this study includes 486 SNPs from 33 genes and these genes along with the distribution of the number of SNPs within each gene is depicted in Figure 1 of the supplementary material. The freely available software package PLINK (Purcell et.al., 2007) was used for genomic quality control.  Thresholds used for SNP and subject exclusion were the same as in Wang et. al. (2012), with the following exceptions. For SNPs, we required a more conservative genotyping call rate of at least 95\% (Ge et al. 2012). 

For subjects, we required at least one baseline and one follow-up MRI scan and excluded multivariate outliers. Sporadically missing genotypes at SNPs in the HapMap3 reference panel (Gibbs et. al., 2003) were imputed into the data using IMPUTE2 (Howie et. al., 2009).  Further details of the quality control and imputation procedure can be found in Szefer (2014). The MRI data from the ADNI-1 database are preprocessed using the FreeSurfer V4 software which conducts automated parcellation to define volumetric and cortical thickness values from the $c=56$ brain regions of interest that are detailed in Table 2. Each of the response variables are adjusted for age, gender, education, handedness, and baseline total intracranial volume (ICV) based on regression weights from healthy controls and are then scaled and centered to have zero-sample-mean and unit-sample-variance. 

We fit our model, which for the current dataset has 27,216 regression parameters, by running a total of $49$ Gibbs sampling chains in parallel on a computing cluster with each chain corresponding to a different value of $(\lambda_{1}^{2},\lambda_{2}^{2})$, and the WAIC is applied to select which of the $49$ chains to use for posterior inference. The Wang et al. (2012) estimator is also computed with tuning parameters $\gamma_{1}$ and $\gamma_{2}$ in (\ref{Wang estimator}) set based on $\gamma_{1} = 2\sigma \lambda_{1}$ and $\gamma_{2} = 2  \sigma \lambda_{2}$, where the values of $\lambda_{1}$ and $\lambda_{2}$ chosen using WAIC and the posterior mean for $\sigma$ from the Gibbs sampler are used.

To select potentially important SNPs we evaluate the 95\% equal-tail credible interval for each regression coefficient and select those SNPs where at least one of the associated credible intervals excludes 0. In total there are 45 SNPs and 152 regression coefficients for which this occurs. Table 1 in the supplementary material lists each of the 152 SNP-ROI associations along with the corresponding point and interval estimates.

The 45 selected SNPs and the corresponding phenotypes at which we see a potential association based on the 95\% credible interval are listed in Table 3. Three SNPs, rs4311 from the ACE gene, rs405509 from the APOE gene, and rs10787010 from the SORCS1 gene stand out as being potentially associated with the largest number of ROIs. The 95\% credible intervals for the coefficients relating rs4311 to each of the $c=56$ imaging measures are depicted in Figure 3, while similar figures for rs405509 and rs10787010 are presented in Figure 2 and Figure 3 of the supplementary material.
\begin{table*}[htbp]
\begin{center}
\tiny
\caption{
The 45 SNPs selected from the Bayesian model along with corresponding phenotypes where (L), (R), (L,R) denote that the phenotypes are on the left, right, and both hemispheres respectively. SNPs also ranked among the top 45 using the Wang et al. (2012) estimate are listed in bold.} 
\begin{tabular}{lll}
\hline
SNP & Gene & Phenotype ID (Hemisphere)\\
\hline
rs4305 &ACE& LatVent (R)\\

{\bf rs4311}&ACE&InfParietal (L,R), MeanPar (L,R), Precuneus (L,R), SupParietal (L), SupTemporal (L), CerebCtx (R),MeanFront (R), \\ 
&&MeanSensMotor (R), MeanTemp (R), Postcentral (R), PostCing (R), Precentral (R), SupFrontal (R), SupParietal (R)\\

{\bf rs405509} & APOE & AmygVol (L), CerebWM (L), Fusiform (L), HippVol (L), InfParietal (L,R),SupFrontal (L,R), Supramarg (L,R), \\ 
&&InfTemporal (L), MeanFront (L,R), MeanLatTemp (L,R), MeanMedTemp (L,R), MeanPar (L,R),\\
&&MeanSensMotor (L,R), MeanTemp (L,R), MidTemporal (L,R), Postcentral (L,R), Precuneus (L,R)\\
&&SupTemporal (L,R), Precentral (R), SupParietal (R)\\

rs11191692 &	CALHM1&	EntCtx (L)\\

{\bf rs3811450} &	CHRNB2&	Precuneus (R)\\

rs9314349 &	CLU&	Parahipp (L)\\

{\bf rs2025935} &	CR1 & CerebWM (R), Fusiform (R), InfLatVent (R)\\

rs11141918 &	DAPK1&	CerebCtx (R) \\

{\bf rs1473180} &	DAPK1 & CerebCtx (L,R) ,EntCtx (L), Fusiform (L), MeanMedTemp (L), MeanTemp (L), PostCing (L)\\

{\bf rs17399090} &	DAPK1 & MeanCing (R), PostCing (R)\\

rs3095747 &	DAPK1 &	InfLatVent (R)\\

{\bf rs3118846} &	DAPK1 &	InfParietal (R)\\

{\bf rs3124237} &	DAPK1 & PostCing (R), Precuneus (R), SupFrontal (R)\\

rs4878117	 & DAPK1 & MeanSensMotor (R), Postcentral (R)\\

rs212539	& ECE1 &	PostCing (R)\\

rs6584307 &	ENTPD7 & Parahipp (L)\\

rs11601726 &	GAB2 & CerebWM (L), LatVent (L)\\

{\bf rs16924159}&	IL33& MeanCing (L), PostCing (L), CerebWM (R)\\

rs928413	& IL33 &	InfLatVent (R)\\

{\bf rs1433099} &	LDLR& CerebCtx.adj (L), Precuneus (L,R)\\

rs2569537 &	LDLR& CerebWM (L,R)\\

rs12209631 &	NEDD9& CerebCtx (L), HippVol (L,R)\\

rs1475345 &	NEDD9 &	Parahipp (L)\\

{\bf rs17496723}&	NEDD9 &	Supramarg (L)\\

rs2327389 &	NEDD9 &	AmygVol (L) \\

{\bf rs744970} &	NEDD9 &	MeanFront (L), SupFrontal (L)\\

{\bf rs7938033} &	PICALM & EntCtx (R), HippVol (R)\\

{\bf rs2756271} &	PRNP & EntCtx (L), HippVol (L,R), InfTemporal (L), Parahipp (L)\\

{\bf rs6107516} &	PRNP & MidTemporal (L,R)\\

rs1023024 &	SORCS1 & MeanSensMotor (L), Precentral (L)\\

{\bf rs10787010} &	SORCS1 & AmygVol (L), EntCtx (L,R), Fusiform (L), HippVol (L,R), InfLatVent (L), InfTemporal (L), MeanFront (L),\\ 
&&MeanMedTemp (L,R), MeanTemp (L), Precentral (L), TemporalPole (R)\\

rs10787011 &	SORCS1& EntCtx (L,R), HippVol(R)\\

rs12248379 &	SORCS1	& PostCing (R)\\

rs1269918 &	SORCS1 &CerebCtx (L), CerebWM (L), InfLatVent (L)\\

{\bf rs1556758} &	SORCS1	&SupParietal (L) \\

{\bf rs2149196} &	SORCS1 & MeanSensMotor (L), Postcentral (L,R)\\

{\bf rs2418811} &	SORCS1 & CerebWM (L,R), InfLatVent.adj (L)\\

{\bf rs10502262} &	SORL1 & MeanCing (L), InfTemporal (R), Supramarg (R)\\

{\bf rs1699102} &	SORL1 & MeanMedTemp (R),  MeanTemp (R)\\

rs1699105 &	SORL1 & MeanCing (L), Precuneus (L)\\

rs4935774&	SORL1	&CerebWM (L,R)\\

rs666004&	SORL1&	InfTemporal (L)\\

rs1568400&	THRA& Precentral (L), TemporalPole (R)\\

rs3744805&	THRA& MeanSensMotor (R), Postcentral (R), Precentral (R)\\

rs7219773&	TNK1& MeanSensMotor (L), Precentral (L), Postcentral (R)\\
\hline
\end{tabular}
\end{center}
\label{default}
\end{table*}%
\begin{figure}[t]
\centering
\includegraphics[scale=0.44]{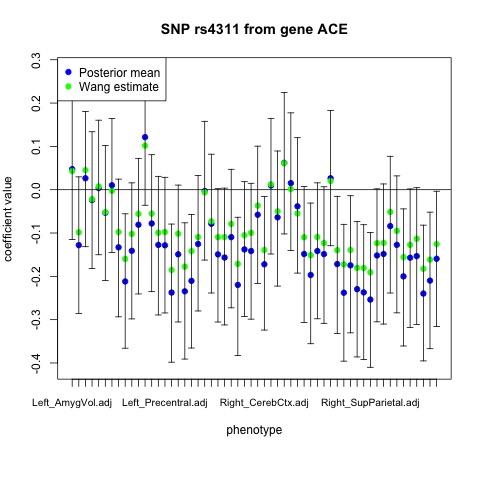}
\caption{The 95\% equal-tail credible intervals relating the SNP rs4311 from ACE to each of the $c=56$ imaging phenotypes, represented on the x-axis in the same order as they are listed in the rows of Table 2, first for the left hemisphere followed by the right.}
\end{figure}

In the original methodology of Wang et al. (2012) the authors suggest ranking and selecting SNPs by constructing a SNP weight based on the point estimate $\hat{\vW}$ and a sum of the absolute values of the estimated coefficients of each single SNP over all of the tasks. Doing so, the top 45 highest ranked SNPs contains 21 of the SNPs chosen using our approach and these 21 SNPs are highlighted in Table 3. The number 1 ranked (highest priority)  SNP using this approach is SNP rs3026841 from gene ECE1. In Figure 4 we display the corresponding point estimates along with the 95\% credible intervals (obtained via our Gibbs sampler) relating this SNP to each of the $c=56$ imaging measures. {\bf We note that all 56 of the corresponding 95\% credible intervals include the value 0.} This result demonstrates clearly the importance of accounting for posterior uncertainty beyond the point estimate and illustrates the potential problems that may arise when estimation uncertainty is ignored. It thus serves to illustrate the practical value of our proposed methodology.
\begin{figure}[t]
\centering
\includegraphics[scale=0.44]{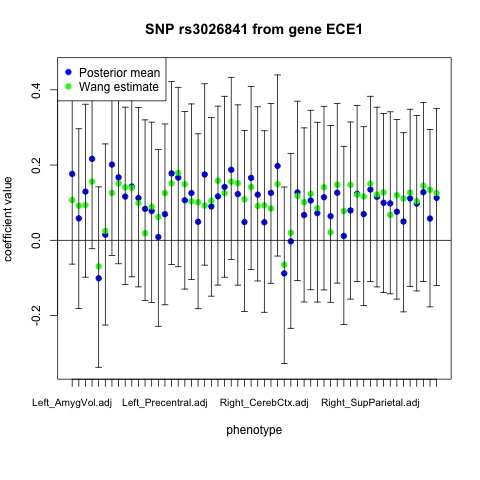}
\caption{The 95\% equal-tail credible intervals relating the SNP rs3026841 from ECE1 to each of the $c=56$ imaging phenotypes represented on the x-axis in the same order as they are listed in the rows of Table 2, first for the left hemisphere followed by the right.}
\end{figure}
\vspace{-0.5cm}

\section{Conclusion}
We have proposed a framework for the analysis of data arising in studies of imaging genomics that extends a previously developed regularization approach in order to allow for the quantification of estimation (posterior) uncertainty in multi-task regression with a $G_{2,1}-norm$ penalty. The value added of our approach has been demonstrated using both simulation studies as well as the analysis of a real dataset from the ADNI database. We have compared our approach to the nonparametric bootstrap applied to (\ref{Wang estimator}) and have demonstrated that our methodology clearly outperforms the latter in terms of mean coverage probability, for the settings considered. We note that our implementation of the bootstrap estimates the tuning parameters from the dataset using cross-validation and subsequently these parameters are fixed across all bootstrap replicates. To keep the computational burden down, it is routine to fix tuning parameters when bootstrapping; however, fixing these parameters does ignore the uncertainty associated with the estimated tuning parameters and this may be contributing to the bias towards below-nominal coverage in the bootstrap intervals. Re-estimating the tuning parameters for each bootstrap replicate is computationally infeasible without massively parallel computers.

It should be noted that we have not addressed statistical adjustments for multiplicity; however, our contribution is a step forward in moving from point estimation to posterior distributions for this regression model. Bayesian false discovery rate procedures (Morris et al., 2008) can be used to adjust for multiplicity in the selection of SNPs based on the output of the Gibbs sampler and this will be considered in future work.

We are currently investigating an extension of the model that allows for a more flexible covariance structure in the specification (\ref{model - level 1}), and alternative shrinkage prior formulations such as the horseshoe prior (Carvalho et al., 2010) that could potentially be further developed for the type of bi-level penalization we have considered here. An alternative approach that is potentially of interest in allowing for increased scalability of the proposed model is the use of a low-rank approximation to the regression coefficient matrix $\vW$ as considered in Marttinen et al. (2014), though this would require an appropriate choice for the rank of the regression model. The sparsity structure we propose in this article could then be incorporated into such an approximation as an extension to the current approach. In addition, extending our model to accommodate potential hidden confounding factors through a joint modelling approach as considered in Fusi et al. (2012), and the incorporation of terms allowing for gene-gene interactions are interesting avenues for future work.

%
%

\section*{Acknowledgements}
Research is supported by funding from the Natural Sciences and Engineering Research Council of Canada. F.S. Nathoo holds a Tier II Canada Research Chair in Biostatistics for Spatial and High-Dimensional Data. Research was enabled in part by support provided by WestGrid (www.westgrid.ca) and Compute Canada (www.computecanada.ca) with assistance provided by HPC specialist Dr. Belaid Moa.  Data collection and sharing for this project was funded by the Alzheimer's Disease Neuroimaging Initiative (ADNI) (National Institutes of Health Grant U01 AG024904) and DOD ADNI (Department of Defense award number W81XWH-12-2-0012). The authors thank Dr. Faisal Beg and Donghuan Lu for assistance with preprocessing of the ADNI MRI data. This work was based on Keelin Greenlaw's MSc thesis supervised by F.S. Nathoo and M. Lesperance.

\bibliographystyle{plain}
%
%

\nocite{*}
\bibliography{main_paper}

\pagebreak

\begin{center}
\textbf{\Large Supplemental Materials: 'A Bayesian Group Sparse Multi-Task Regression Model for Imaging Genetics'}
\end{center}

\setcounter{equation}{0}
\setcounter{figure}{0}
\setcounter{table}{0}
\setcounter{page}{1}
\setcounter{section}{0}
\makeatletter
\renewcommand{\theequation}{S\arabic{equation}}
\renewcommand{\thefigure}{S\arabic{figure}}
\renewcommand{\thetable}{S\arabic{table}}


\def \veta{\mbox{\boldmath $\eta$ \unboldmath}\!\!}
\def \vnu{\mbox{\boldmath $\nu$ \unboldmath}\!\!}
\def \ve{\mbox{\boldmath $e$ \unboldmath}\!\!}
\def \vZ{\mbox{\boldmath $Z$ \unboldmath}\!\!}
\def \vR{\mbox{\boldmath $R$ \unboldmath}\!\!}
\def \vI{\mbox{\boldmath $I$ \unboldmath}\!\!}
\def \vW{\mbox{\boldmath $W$ \unboldmath}\!\!}
\def \vB{\mbox{\boldmath $B$ \unboldmath}\!\!}
\def \vF{\mbox{\boldmath $F$ \unboldmath}\!\!}
\def \vp{\mbox{\boldmath $p$ \unboldmath}\!\!}
\def \vU{\mbox{\boldmath $U$ \unboldmath}\!\!}
\def \vV{\mbox{\boldmath $V$ \unboldmath}\!\!}
\def \vD{\mbox{\boldmath $D$ \unboldmath}\!\!}
\def \vSigma{\mbox{\boldmath $\Sigma$ \unboldmath}\!\!}
\def \vTheta{\mbox{\boldmath $\Theta$ \unboldmath}\!\!}
\def \vLambda{\mbox{\boldmath $\Lambda$ \unboldmath}\!\!}
\def \vomega{\mbox{\boldmath $\omega$ \unboldmath}\!\!}
\def \vbeta{\mbox{\boldmath $\beta$ \unboldmath}\!\!}
\def \veta{\mbox{\boldmath $\eta$ \unboldmath}\!\!}
\def \valpha{\mbox{\boldmath $\alpha$ \unboldmath}\!\!}
\def \vgamma{\mbox{\boldmath $\gamma$ \unboldmath}\!\!}
\def \vepsilon{\mbox{\boldmath $\epsilon$ \unboldmath}\!\!}
\def \vrho{\mbox{\boldmath $\rho$ \unboldmath}\!\!}
\def \va{\mbox{\boldmath $a$ \unboldmath}\!\!}
\def \vs{\mbox{\boldmath $s$ \unboldmath}\!\!}
\def \vai{\mbox{\boldmath $a^{(i)}$ \unboldmath}\!\!}
\def \vu{\mbox{\boldmath $u$ \unboldmath}\!\!}
\def \vh{\mbox{\boldmath $h$ \unboldmath}\!\!}
\def \vb{\mbox{\boldmath $b$ \unboldmath}\!\!}
\def \vbs{\mbox{\boldmath $b(s)$ \unboldmath}\!\!}
\def \vbsi{\mbox{\boldmath $b(s_{i})$ \unboldmath}\!\!}
\def \vZ{\mbox{\boldmath $Z$ \unboldmath}\!\!}
\def \vA{\mbox{\boldmath $A$ \unboldmath}\!\!}
\def \vZss{\mbox{\boldmath $z^{*}(s)$ \unboldmath}\!\!}
\def \vZsi{\mbox{\boldmath $z(s_{i})$ \unboldmath}\!\!}
\def \vt{\mbox{\boldmath $t$ \unboldmath}\!\!}
\def \vM{\mbox{\boldmath $M$ \unboldmath}\!\!}
\def \vE{\mbox{\boldmath $E$ \unboldmath}\!\!}
\def \vTheta{\mbox{\boldmath $\Theta$ \unboldmath}\!\!}
\def \vtau{\mbox{\boldmath $\tau$ \unboldmath}\!\!}
\def \vomega{\mbox{\boldmath $\omega$ \unboldmath}\!\!}
\def \vmu{\mbox{\boldmath $\mu$ \unboldmath}\!\!}
\def \vmuZ{\mbox{\boldmath $\mu_{z}$ \unboldmath}\!\!}
\def \vpsi{\mbox{\boldmath $\psi$ \unboldmath}\!\!}
\def \vK{\mbox{\boldmath $K$ \unboldmath}\!\!}
\def \vY{\mbox{\boldmath $Y$ \unboldmath}\!\!}
\def \vL{\mbox{\boldmath $L$ \unboldmath}\!\!}
\def \vS{\mbox{\boldmath $S$ \unboldmath}\!\!}
\def \vx{\mbox{\boldmath $x$ \unboldmath}\!\!}
\def \vX{\mbox{\boldmath $X$ \unboldmath}\!\!}
\def \vWu{\mbox{\boldmath $w(u)$ \unboldmath}\!\!}

\section{Derivations for the Gibbs Sampling Algorithm}

Here we derive the full conditional distributions required for Gibbs sampling. In what follows we make use of the vectorisation of a matrix, $\boldsymbol{A}$, a linear transformation of $\boldsymbol{A}$ to a column vector in which the columns of $\boldsymbol{A}$ are stacked one under the other to form a single column. Eg. If  $\boldsymbol{A}$ is a $d \times c$ matrix, then $\text{vec}(\boldsymbol{A}^T) = [ A_{1,1}, \dots , A_{1,c}, A_{2,1}, \dots, A_{2,c}, \dots, A_{d,1}, \dots, A_{d,c} ] ^T.$ We will further make use of the result $\text{vec}(\boldsymbol{A}\boldsymbol{B}) = (\boldsymbol{B}^T \otimes I_k)\text{vec}(\boldsymbol{A})$, where $k$ is the number of rows in $A$ and $\otimes$ denotes the Kronecker product. 

Assuming the scale mixture representation of the product multivariate Laplace distribution presented in Proposition 2, the joint posterior distribution can be expressed up to a normalizing constant as

\begin{equation*}
\begin{split}
&     p  (\boldsymbol{W},  \tauo, \cdots, \tauK, \omegao, \cdots, \omegad, \sig | \boldsymbol{Y} ) {}  \propto{}  p(\boldsymbol{Y} | \boldsymbol{W}, \sig) p(\boldsymbol{W} | \sig, \vtau^2,  \vomega)p( \vtau^{2}, \vomega^{2} |  \lamo, \lamt)p(\sig | a_{\sigma}, b_{\sigma} )\\
&\propto{}  |\sig I_c |^{-\frac{n}{2}} \exp \left\lbrace - \frac{1}{2\sig}\sum_{\ell=1}^n ( \boldsymbol{y}_{\ell} - \boldsymbol{W}^T\boldsymbol{x}_{\ell})^T  (\boldsymbol{y}_{\ell} - \boldsymbol{W}^T\boldsymbol{x}_{\ell}) \right\rbrace \\
& \times \prodkK \left[  \prod_{i \in \pi_k} \left[ \left(\sig \left(\frac{1}{\tau_{k}^2}\:+\:\frac{1}{\omega_i^2}\right)^{-1}\right)^{-\frac{c}{2}} \right] \exp \left\lbrace - \sumink \left( \frac{  \sumjc w_{ij}^2}{2\sig \left(\frac{1}{\tau_{k}^2}\:+\:\frac{1}{\omega_i^2}\right)^{-1}} \right) \right\rbrace \right] \\
&\times \prodkK  \left(\frac{\lambda_1^2}{2}\right)^{\left(\frac{m_kc + 1}{2}\right)}(\tau_k^2)^{\left(\frac{m_kc + 1}{2}\right) -1 } \exp \left\lbrace -\left(\frac{\lambda_1^2}{2}\right) \tau_k^2 \right\rbrace\\
&\times \left[ \prod_{i \in \pi_{k}}   \left(\frac{\lambda_2^2}{2}\right)^{\left(\frac{c + 1}{2}\right)}(\omega_i^{2})^{\left(\frac{c + 1}{2}\right) -1 } \exp \left\lbrace -\left(\frac{\lambda_2^2}{2}\right) \omega_{i}^{2} \right\rbrace (\tau_{k}^{2}+ \omega_{i}^{2})^{-\frac{c}{2}}\right]\\
&\times \frac{b_{\sigma}^{a_{\sigma}}}{\Gamma (a_{\sigma})} (\sig)^{-a_{\sigma} -1} \exp \left\lbrace - \frac{b_{\sigma}}{\sig} \right\rbrace. 
\end{split}
\end{equation*}

\bigskip

\noindent{\bf{Full conditional distribution of $\boldsymbol{W^{(k)}}$}}
\begin{multline}  \label{wkfull1}
p(\boldsymbol{W}^{(k)} \big| \boldsymbol{Y}, \boldsymbol{W}^{(-k)}, \vtau^{2}, \vomega^{2}, \sig, \lamo, \lamt) \propto  \\
\exp \left\lbrace \frac{-1}{2\sig} \sum_{\ell=1}^n ( \boldsymbol{y}_{\ell} - \boldsymbol{W}^T\boldsymbol{x}_{\ell})^T  (\boldsymbol{y}_{\ell} - \boldsymbol{W}^T\boldsymbol{x}_{\ell}) \right\rbrace \exp \left\lbrace - \sumink \left( \frac{ \sumjc w_{ij}^2}{2\sig\left(\frac{1}{\tau_{k}^2}\:+\:\frac{1}{\omega_i^2}\right)^{-1}} \right) \right\rbrace .
\end{multline}

Split $\boldsymbol{W}$ into $\boldsymbol{W}^{(k)}$ and $\boldsymbol{W}^{(-k)}$ and rewrite the first exponent of (\ref{wkfull1}) as
\begin{equation} \label{wkfull2}
\exp \left\lbrace \frac{-1}{2\sig} \sumln (\boldsymbol{y}_{\ell} - \boldsymbol{W}^{(k) T}\boldsymbol{x}_{\ell}^{(k)} - \boldsymbol{W}^{(-k)T}\boldsymbol{x}_{\ell}^{(-k)})^T(\boldsymbol{y}_{\ell} - \boldsymbol{W}^{(k) T}\boldsymbol{x}_{\ell}^{(k)} - \boldsymbol{W}^{(-k)T}\boldsymbol{x}_{\ell}^{(-k)}) \right\rbrace. 
\end{equation}

We vectorise the terms that include either $\boldsymbol{W}^{(k)}$ or $\boldsymbol{W}^{(-k)}$ and simplify based on:

1) $\text{vec}(\boldsymbol{W}^{(k) T}\boldsymbol{x}_{\ell}^{(k)}) = (\boldsymbol{x}_{\ell}^{(k)T} \otimes I_c)\text{vec}(\boldsymbol{W}^{(k) T})$

2) $\text{vec}(\boldsymbol{W}^{(-k)T}\boldsymbol{x}_{\ell}^{(-k)}) = (\boldsymbol{x}_{\ell}^{(-k)T} \otimes I_c) \text{vec}(\boldsymbol{W}^{(-k)T}).$

\medskip

These results give an equivalent expression for (\ref{wkfull2})
\begin{multline*}
\exp \left\lbrace \frac{-1}{2\sig} \sumln \left(\boldsymbol{y}_{\ell} - (\boldsymbol{x}_{\ell}^{(k)T} \otimes I_c)\text{vec}(\boldsymbol{W}^{(k) T}) - (\boldsymbol{x}_{\ell}^{(-k)T} \otimes I_c) \text{vec}(\boldsymbol{W}^{(-k)T})\right)^T  \right. \\ 
\left. \left(\boldsymbol{y}_{\ell} - (\boldsymbol{x}_{\ell}^{(k)T} \otimes I_c)\text{vec}(\boldsymbol{W}^{(k) T}) - (\boldsymbol{x}_{\ell}^{(-k)T} \otimes I_c) \text{vec}(\boldsymbol{W}^{(-k)T})\right) \right\rbrace 
\end{multline*}
which can be expressed as
\begin{multline*}
\exp \left\lbrace \frac{-1}{2\sig} \sumln \left(\boldsymbol{y}_{\ell}^T - \text{vec}(\boldsymbol{W}^{(k) T})^T(\boldsymbol{x}_{\ell}^{(k)T} \otimes I_c)^T - \text{vec}(\boldsymbol{W}^{(-k)T})^T(\boldsymbol{x}_{\ell}^{(-k)T} \otimes I_c)^T \right)  \right. \\
\left.\left(\boldsymbol{y}_{\ell} - (\boldsymbol{x}_{\ell}^{(k)T} \otimes I_c)\text{vec}(\boldsymbol{W}^{(k) T}) - (\boldsymbol{x}_{\ell}^{(-k)T} \otimes I_c) \text{vec}(\boldsymbol{W}^{(-k)T})\right) \right\rbrace 
\end{multline*}

Using $(A \otimes B)^T = (A^T \otimes B^T)$ the above is simplified to 
\begin{multline*}
\exp \left\lbrace \frac{-1}{2\sig} \sumln \left(\boldsymbol{y}_{\ell}^T - \text{vec}(\boldsymbol{W}^{(k) T})^T(\boldsymbol{x}_{\ell}^{(k)} \otimes I_c) - \text{vec}(\boldsymbol{W}^{(-k)T})^T(\boldsymbol{x}_{\ell}^{(-k)} \otimes I_c) \right) \right. \\ 
\left. \left(\boldsymbol{y}_{\ell} - (\boldsymbol{x}_{\ell}^{(k)T} \otimes I_c)\text{vec}(\boldsymbol{W}^{(k) T}) - (\boldsymbol{x}_{\ell}^{(-k)T} \otimes I_c) \text{vec}(\boldsymbol{W}^{(-k)T})\right) \right\rbrace. 
\end{multline*}
It is now possible to expand the expression. Only those terms that include $\boldsymbol{W}^{(k)}$ are kept, as the other terms are considered to be constants that can be factored out to become part of the normalising constant. We have
\begin{multline*}
\exp  \left\lbrace  \frac{-1}{2\sig} \sumln  \left( -\boldsymbol{y}_{\ell}^T (\boldsymbol{x}_{\ell}^{(k)T} \otimes I_c) \text{vec}(\boldsymbol{W}^{(k) T}) -\text{vec}(\boldsymbol{W}^{(k) T})^T (\boldsymbol{x}_{\ell}^{(k)} \otimes I_c) \boldsymbol{y}_{\ell} \right.\right. \\
\left.\left. + \text{vec}(\boldsymbol{W}^{(k) T})^T (\boldsymbol{x}_{\ell}^{(k)} \otimes I_c)(\boldsymbol{x}_{\ell}^{(k)T} \otimes I_c) \text{vec}(\boldsymbol{W}^{(k) T}) \right.\right. \\ \left.\left. + \text{vec}(\boldsymbol{W}^{(k) T})^T (\boldsymbol{x}_{\ell}^{(k)} \otimes I_c)(\boldsymbol{x}_{\ell}^{(-k)T} \otimes I_c) \text{vec}(\boldsymbol{W}^{(-k)T}) \right. \right. \\
+ \left.\left.\text{vec}(\boldsymbol{W}^{(-k)T})^T (\boldsymbol{x}_{\ell}^{(-k)} \otimes I_c) (\boldsymbol{x}_{\ell}^{(k)T} \otimes I_c) \text{vec}(\boldsymbol{W}^{(k) T}) \right)   \right\rbrace 
\end{multline*}
which can be expressed as
\begin{multline*}
\exp \left\lbrace  \frac{-1}{2\sig} \left[ \text{vec}(\boldsymbol{W}^{(k) T})^T \sumln (\boldsymbol{x}_{\ell}^{(k)} \otimes I_c)(\boldsymbol{x}_{\ell}^{(k)T} \otimes I_c) \text{vec}(\boldsymbol{W}^{(k) T})  \right.\right. \\ \left. \left. + 2 \text{vec}(\boldsymbol{W}^{(k) T})^T \sumln (\boldsymbol{x}_{\ell}^{(k)} \otimes I_c)(\boldsymbol{x}_{\ell}^{(-k)T} \otimes I_c) \text{vec}(\boldsymbol{W}^{(-k)T})  - 2 \text{vec}(\boldsymbol{W}^{(k) T})^T \sumln (\boldsymbol{x}_{\ell}^{(k)} \otimes I_c) \boldsymbol{y}_{\ell} \right] \right\rbrace. 
\end{multline*}
Next, consider the second exponent in (\ref{wkfull1}), 

$$\exp \left\lbrace \frac{-1}{2\sig} \sumink  \frac{  \sumjc w_{ij}^2}{\left(\frac{1}{\tau_{k}^2}\:+\:\frac{1}{\omega_i^2}\right)^{-1}},  \right\rbrace$$

and define a matrix, $\boldsymbol{H}_k$, such that $ \boldsymbol{H}_k = \left[ diag \left\lbrace \frac{1}{\tauk} + \frac{1}{\omegai} \right\rbrace_{i \in \pi_k} \otimes I_c \right].$

Notice that, 
$$ \sumink  \frac{  \sumjc w_{ij}^2}{\left(\frac{1}{\tau_{k}^2}\:+\:\frac{1}{\omega_i^2}\right)^{-1}}  =   \text{vec}(\boldsymbol{W}^{(k) T})^T \boldsymbol{H}_k \text{vec}(\boldsymbol{W}^{(k) T}) .$$

We can then rewrite (\ref{wkfull1}), up to its normalising constant, as,  
\begin{multline} \label{wkfull3}
p(\boldsymbol{W}^{(k)} \big| \boldsymbol{Y}, \boldsymbol{W}^{(-k)}, \vtau, \vomega, \sig, \lamo, \lamt) \propto  \\
\exp \left\lbrace  \frac{-1}{2\sig} \left[ \text{vec}(\boldsymbol{W}^{(k) T})^T \sumln (\boldsymbol{x}_{\ell}^{(k)} \otimes I_c)(\boldsymbol{x}_{\ell}^{(k)T} \otimes I_c) \text{vec}(\boldsymbol{W}^{(k) T}) \right.\right. \left.\left. + \text{vec}(\boldsymbol{W}^{(k) T})^T \boldsymbol{H}_k \text{vec}(\boldsymbol{W}^{(k) T})  \right.\right. \\ \left.\left. + 2 \text{vec}(\boldsymbol{W}^{(k) T})^T \sumln (\boldsymbol{x}_{\ell}^{(k)} \otimes I_c)(\boldsymbol{x}_{\ell}^{(-k)T} \otimes I_c) \text{vec}(\boldsymbol{W}^{(-k)T}) \right.\right. \\  \left.\left. - 2 \text{vec}(\boldsymbol{W}^{(k) T})^T \sumln (\boldsymbol{x}_{\ell}^{(k)} \otimes I_c) \boldsymbol{y}_{\ell} \right] \right\rbrace. 
\end{multline}
Expression (\ref{wkfull3}) is a quadratic form in $\text{vec}(\boldsymbol{W}^{(k) T})$ in the exponent. Therefore, the full conditional distribution of $\text{vec}(\boldsymbol{W}^{(k) T})$ is multivariate normal of dimension $m_kc$, with parameters, say, $\vmuk$ and $\boldsymbol{\Sigma}_k$.
After expanding, the exponent of a multivariate normal distribution is of the form, \begin{equation} \label{eq:mvn}
\exp \left\lbrace -\frac{1}{2} \left[ \text{vec}(\boldsymbol{W}^{(k) T})^T \boldsymbol{\Sigma}_k^{-1} \text{vec}(\boldsymbol{W}^{(k) T})-2 \text{vec}(\boldsymbol{W}^{(k) T})^T \boldsymbol{\Sigma}_k^{-1} \vmuk + \;constant\; \right]\right\rbrace . \end{equation}
The next steps involve matching (\ref{wkfull3}) to (\ref{eq:mvn}). 

\noindent \emph{Solving for $\boldsymbol{\Sigma}_k$:}\\
Consider the terms of (\ref{wkfull3}) that are quadratic in $\text{vec}(\boldsymbol{W}^{(k) T})$. We have $$ \exp \left\lbrace  \frac{-1}{2\sig} \left[ \text{vec}(\boldsymbol{W}^{(k) T})^T \sumln (\boldsymbol{x}_{\ell}^{(k)} \otimes I_c)(\boldsymbol{x}_{\ell}^{(k)T} \otimes I_c) \text{vec}(\boldsymbol{W}^{(k) T}) +
\text{vec}(\boldsymbol{W}^{(k) T})^T \boldsymbol{H}_k \text{vec}(\boldsymbol{W}^{(k) T}) \right] \right\rbrace.$$
Rearrange to obtain
$$ \exp \left\lbrace  -\frac{1}{2} \left[ \text{vec}(\boldsymbol{W}^{(k) T})^T \left(   \frac{1}{\sig} \left( \sumln (\boldsymbol{x}_{\ell}^{(k)} \otimes I_c)(\boldsymbol{x}_{\ell}^{(k)T} \otimes I_c) +  \boldsymbol{H}_k \right) \right) \text{vec}(\boldsymbol{W}^{(k) T}) \right] \right\rbrace$$
We now observe that 
\begin{align*}
\boldsymbol{\Sigma}_k^{-1} = &   \frac{1}{\sig} \left( \sumln (\boldsymbol{x}_{\ell}^{(k)} \otimes I_c)(\boldsymbol{x}_{\ell}^{(k)T} \otimes I_c) + \boldsymbol{H}_k \right),  \\
\boldsymbol{\Sigma}_k = &  \sig \left( \sumln (\boldsymbol{x}_{\ell}^{(k)} \otimes I_c)(\boldsymbol{x}_{\ell}^{(k)T} \otimes I_c) + \boldsymbol{H}_k \right)^{-1}. 
\end{align*}
This gives $\boldsymbol{\Sigma}_k = \sig \boldsymbol{A}_k^{-1}$,\\  where $\boldsymbol{A}_k = \left( \sumln (\boldsymbol{x}_{\ell}^{(k)} \otimes I_c)(\boldsymbol{x}_{\ell}^{(k)T} \otimes I_c) + \left( diag \left\lbrace \frac{1}{\tauk} + \frac{1}{\omegai} \right\rbrace_{i \in \pi_k} \otimes I_c \right) \right). $

\noindent \emph{Solving for $\vmuk$}:\\
Consider the term $- \frac{1}{2} \left( -2 \text{vec}(\boldsymbol{W}^{(k) T})^T \boldsymbol{\Sigma}_k^{-1} \vmuk \right)$ within the density of the multivariate normal density. We have the expression,
\begin{multline*}
- \frac{1}{2\sig}  \left( 2 \text{vec}(\boldsymbol{W}^{(k) T})^T \sumln (\boldsymbol{x}_{\ell}^{(k)} \otimes I_c)(\boldsymbol{x}_{\ell}^{(-k)T} \otimes I_c) \text{vec}(\boldsymbol{W}^{(-k)T})  - 2 \text{vec}(\boldsymbol{W}^{(k) T})^T \sumln (\boldsymbol{x}_{\ell}^{(k)} \otimes I_c) \boldsymbol{y}_{\ell}  \right) \\
=  \text{vec}(\boldsymbol{W}^{(k) T})^T \left( \frac{1}{\sig}\left(- \sumln (\boldsymbol{x}_{\ell}^{(k)} \otimes I_c)(\boldsymbol{x}_{\ell}^{(-k)T} \otimes I_c) \text{vec}(\boldsymbol{W}^{(-k)T}) + \sumln (\boldsymbol{x}_{\ell}^{(k)} \otimes I_c) \boldsymbol{y}_{\ell} \right)  \right).
\end{multline*}

Match up the expressions.
$$\boldsymbol{\Sigma}_k^{-1} \vmuk = \frac{1}{\sig}\left( - \sumln (\boldsymbol{x}_{\ell}^{(k)} \otimes I_c)(\boldsymbol{x}_{\ell}^{(-k)T} \otimes I_c) \text{vec}(\boldsymbol{W}^{(-k)T}) + \sumln (\boldsymbol{x}_{\ell}^{(k)} \otimes I_c) \boldsymbol{y}_{\ell} \right). $$

Isolate $\vmuk$ to obtain
\begin{align*}
\vmuk = & \boldsymbol{\Sigma}_k  \left( \frac{1}{\sig}\left( - \sumln (\boldsymbol{x}_{\ell}^{(k)} \otimes I_c)(\boldsymbol{x}_{\ell}^{(-k)T} \otimes I_c) \text{vec}(\boldsymbol{W}^{(-k)T}) + \sumln (\boldsymbol{x}_{\ell}^{(k)} \otimes I_c) \boldsymbol{y}_{\ell} \right) \right) \\
= & \boldsymbol{A}_k^{-1} \left( - \sumln (\boldsymbol{x}_{\ell}^{(k)} \otimes I_c)(\boldsymbol{x}_{\ell}^{(-k)T} \otimes I_c) \text{vec}(\boldsymbol{W}^{(-k)T}) + \sumln (\boldsymbol{x}_{\ell}^{(k)} \otimes I_c) \boldsymbol{y}_{\ell} \right).
\end{align*}

Finally, the full conditional distribution of $\boldsymbol{W^{(k)}}$ is expressed as $$ \text{vec}(\boldsymbol{W}^{(k) T}) \big| \boldsymbol{Y}, \boldsymbol{W}^{(-k)}, \vtau, \vomega, \sig, \lamo, \lamt \sim MVN_{m_kc} ( \; \vmuk ,\; \boldsymbol{\Sigma}_k ),$$ where $$ \vmuk = \boldsymbol{A}_k^{-1} \left( - \sumln (\boldsymbol{x}_{\ell}^{(k)} \otimes I_c)(\boldsymbol{x}_{\ell}^{(-k)T} \otimes I_c) \text{vec}(\boldsymbol{W}^{(-k)T}) + \sumln (\boldsymbol{x}_{\ell}^{(k)} \otimes I_c) \boldsymbol{y}_{\ell} \right),$$
$$\boldsymbol{A}_k = \left( \sumln (\boldsymbol{x}_{\ell}^{(k)} \otimes I_c)(\boldsymbol{x}_{\ell}^{(k)T} \otimes I_c) + \left( diag \left\lbrace \frac{1}{\tauk} + \frac{1}{\omegai} \right\rbrace_{i \in \pi_k} \otimes I_c \right) \right),\;\; \text{and} \;\; \boldsymbol{\Sigma}_k = \sig \boldsymbol{A}_k^{-1}.$$

\noindent{\emph{Full conditional distribution of $\sig$}}:
\begin{align*}
\begin{split}
p( \sig \big| & \boldsymbol{Y}, \boldsymbol{W}, \vtau, \vomega, \lamo, \lamt)  \\
\propto {} & |\sig I_c|^{-\frac{n}{2}} \exp \left\lbrace - \frac{1}{2\sig} \sum_{\ell=1}^n ( \boldsymbol{y}_{\ell} - \boldsymbol{W}^T\boldsymbol{x}_{\ell})^T  (\boldsymbol{y}_{\ell} - \boldsymbol{W}^T\boldsymbol{x}_{\ell}) \right\rbrace \\ &{} \prodkK \left[ \left( \sig \right)^{-\frac{m_kc}{2}} \prodink \left[ \left(\frac{1}{\tau_{k}^2}\:+\:\frac{1}{\omega_i^2}\right)^{-1} \right]^{-\frac{c}{2}} \exp \left\lbrace - \frac{1}{2 \sig} 
\sumink \frac{ \sumjc w_{ij}^2}{\left(\frac{1}{\tau_{k}^2}\:+\:\frac{1}{\omega_i^2}\right)^{-1}} \right\rbrace \right] \cdot (\sig)^{-a_{\sigma}-1} \exp \left\lbrace -\frac{b_{\sigma}}{\sig} \right\rbrace \\
\\  
={}& \prodkK\prodink \left[ \left(\frac{1}{\tau_{k}^2}\:+\:\frac{1}{\omega_i^2}\right)^{-1} \right]^{-\frac{c}{2}} \left( \sig \right)^{-\frac{cn}{2}} \left( \sig \right)^{-\frac{dc}{2}} \left( \sig \right)^{-a_\sigma -1}\\  
&{} \exp \left\lbrace - \frac{1}{2\sig}  \sum_{\ell=1}^n || \boldsymbol{y}_{\ell} - \boldsymbol{W}^T\boldsymbol{x}_{\ell} ||_2^2 - \frac{1}{2 \sig} 
\sumid \frac{ \sumjc w_{ij}^2}{\left(\frac{1}{\tau_{k(i)}^2}\:+\:\frac{1}{\omega_i^2}\right)^{-1}} -\frac{b_{\sigma}}{\sig}   \right\rbrace.
\end{split}
\end{align*}
Since $\prodkK\prodink \left[ \left(\frac{1}{\tau_{k}^2}\:+\:\frac{1}{\omega_i^2}\right)^{-1} \right]^{-\frac{c}{2}}$ does not depend on $\sig$, it can be factored out of the expression. This step leaves, 
\begin{multline*} 
p(\sig \big| \boldsymbol{Y}, \boldsymbol{W}, \vtau, \vomega, \lamo, \lamt) \propto \\
(\sig) ^{ -\left(\frac{cn}{2} +\frac{dc}{2} +a_{\sigma}\right) -1} \exp \left\lbrace -\frac{1}{\sig} \left( \frac{1}{2} \sum_{\ell=1}^n || \boldsymbol{y}_{\ell} - \boldsymbol{W}^T\boldsymbol{x}_{\ell} ||_2^2  + \frac{1}{2} \sumid \frac{ \sumjc w_{ij}^2}{\left(\frac{1}{\tau_{k(i)}^2}\:+\:\frac{1}{\omega_i^2}\right)^{-1}} +b_{\sigma} \right)  \right\rbrace,
\end{multline*}
so that
\begin{multline*}
\sig \big| \boldsymbol{Y}, \boldsymbol{W}, \vtau, \vomega, \lamo, \lamt \sim \;\; Inv-Gamma \left( a_\sigma^*, b_\sigma^* \right),\\
\text{where}\;\;  a_\sigma^* = \left(\frac{cn}{2} +\frac{dc}{2} +a_{\sigma}\right),  \; b_\sigma^* = \left( \frac{1}{2} \sum_{\ell=1}^n || \boldsymbol{y}_{\ell} - \boldsymbol{W}^T\boldsymbol{x}_{\ell} ||_2^2  + \frac{1}{2} \sumid \frac{ \sumjc w_{ij}^2}{\left(\frac{1}{\tau_{k(i)}^2}\:+\:\frac{1}{\omega_i^2}\right)^{-1}} +b_{\sigma} \right) .
\end{multline*} 

\noindent{\emph{Full Conditional of $\vomega^{2}, \vtau^{2}$}}

We consider a joint update of the scale mixing variable based on the corresponding full conditional distribution. We have $p(\vtau^{2}, \vomega^{2} \big| \boldsymbol{Y}, \boldsymbol{W}, \sig, \lamt, \lamo)$
\begin{align*}
\begin{split}
\propto  & \prodkK \left[  \prod_{i \in \pi_k} \left[ \left(\sig \left(\frac{1}{\tau_{k}^2}\:+\:\frac{1}{\omega_i^2}\right)^{-1}\right)^{-\frac{c}{2}} \right] \exp \left\lbrace - \sumink \left( \frac{  \sumjc w_{ij}^2}{2\sig \left(\frac{1}{\tau_{k}^2}\:+\:\frac{1}{\omega_i^2}\right)^{-1}} \right) \right\rbrace \right] \\
\times&\prodkK  \left(\frac{\lambda_1^2}{2}\right)^{\left(\frac{m_kc + 1}{2}\right)}(\tau_k^2)^{\left(\frac{m_kc + 1}{2}\right) -1 } \exp \left\lbrace -\left(\frac{\lambda_1^2}{2}\right) \tau_k^2 \right\rbrace\\
\times&\left[ \prod_{i \in \pi_{k}}   \left(\frac{\lambda_2^2}{2}\right)^{\left(\frac{c + 1}{2}\right)}(\omega_i^{2})^{\left(\frac{c + 1}{2}\right) -1 } \exp \left\lbrace -\left(\frac{\lambda_2^2}{2}\right) \omega_{i}^{2} \right\rbrace (\tau_{k}^{2}+ \omega_{i}^{2})^{-\frac{c}{2}}\right]. 
\end{split}
\end{align*}
\begin{align*}
\begin{split}
\propto  &  \prodkK (\tau_k^2)^{-\frac{1}{2}} \exp \left\lbrace -\left(\frac{\lambda_1^2}{2}\right)\tau_k^2 - \frac{||\vW^{(k)}||_{2}^{2}}{\tau_k^2 2 \sigma^{2}} \right\rbrace\\
&\times \prodkK  \prod _{i \in \pi_{k}} (\omega_i^2)^{-\frac{1}{2}} \exp \left\lbrace -\left(\frac{\lambda_2^2}{2}\right)\omega_{i}^{2} - \frac{||\boldsymbol{w}^i||_{2}^{2}}{\omega_{i}^{2} 2 \sigma^{2}} \right\rbrace
\end{split}
\end{align*}
where $\boldsymbol{w}^i$ denotes the $i^{th}$ row of $\boldsymbol{W}$. The above expression shows that the scale mixing variables are conditionally independent given $\boldsymbol{Y}, \boldsymbol{W}, \sig, \lamt, \lamo$. We next apply a transformation of variables $\nu_k  = (\tauk)^{-1}$, Jacobian = $\big| \frac{d}{d \nu_k} \tau_k^2 (\nu_k) \big| =  \nu_k^{-2}$;  $\eta_{i}  = (\omega_{i}^{2})^{-1}$, Jacobian = $\big| \frac{d}{d \eta_{i}} \omega_{i}^{2} \big| =  \eta_{i}^{-2}$ which yields $p(\vnu, \veta \big| \boldsymbol{Y}, \boldsymbol{W}, \sig, \lamt, \lamo)$
\begin{align*}
\begin{split}
\propto  &  \prodkK (\nu_k)^{-\frac{3}{2}} \exp \left\lbrace -\left(\frac{\lambda_1^2}{2\nu_{k}}\right) - \frac{\nu_{k}||\vW^{(k)}||_{2}^{2}}{2 \sigma^{2}} \right\rbrace\times \prodkK  \prod _{i \in \pi_{k}} (\eta_{i})^{-\frac{3}{2}} \exp \left\lbrace -\left(\frac{\lambda_2^2}{2 \eta_{i}}\right) 
- \frac{\eta_{i}||\boldsymbol{w}^i||_{2}^{2}}{2 \sigma^{2}} \right\rbrace
\end{split}
\end{align*}
and from this we see that the conditional distributions lie within the Inverse Gaussian family. More specifically we have
$$\nu_k = \frac{1}{\tauk}\;\; \Big|\; \boldsymbol{Y}, \boldsymbol{W}, \sig, \lamo, \lamt \stackrel{ind}{\sim} \textit{Inverse-Gaussian} \left( \sqrt{ \frac{\lamo \sig}{||\boldsymbol{W}^{(k)}||_2^2}}\;,\;\; \lamo \right), \,\,\, k=1,\dots,K $$
independent of
$$\eta_i= \frac{1}{\omegai} \; \;\Big| \; \boldsymbol{Y}, \boldsymbol{W}, \sig, \lamo, \lamt \sim \textit{Inverse-Gaussian} \left( \sqrt{ \frac{\lamt \sig}{||\boldsymbol{w}^i||_2^2}} \;, \;\;\lamt \right), \,\,\, i=1,\dots,d.$$

\section{Supplementary Figures and Tables}

\begin{figure*}[htbp]
\centering
\includegraphics[scale=0.5]{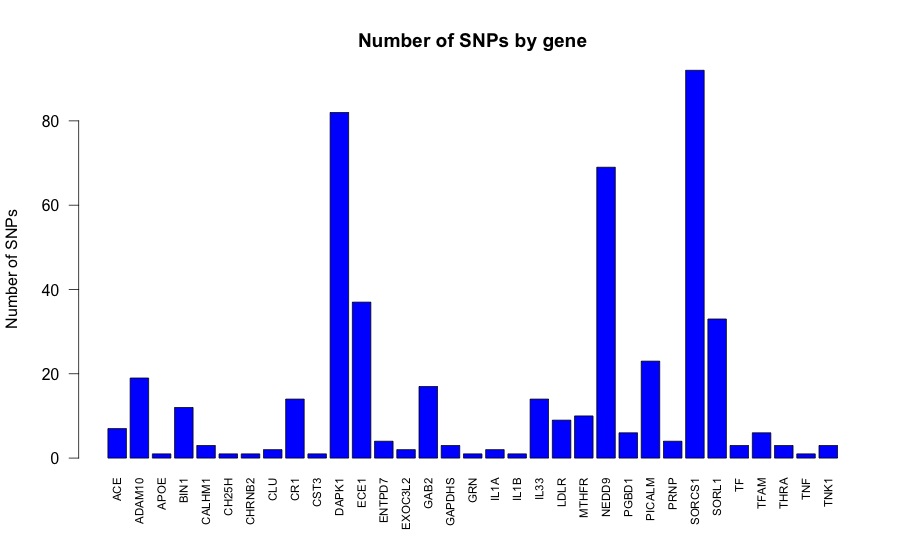}
\caption{Each of the 33 genes partitioning the 486 SNPs included in the simulation studies and data analysis.}
\end{figure*}

\begin{figure}[!t]
\centering
\includegraphics[scale=0.7]{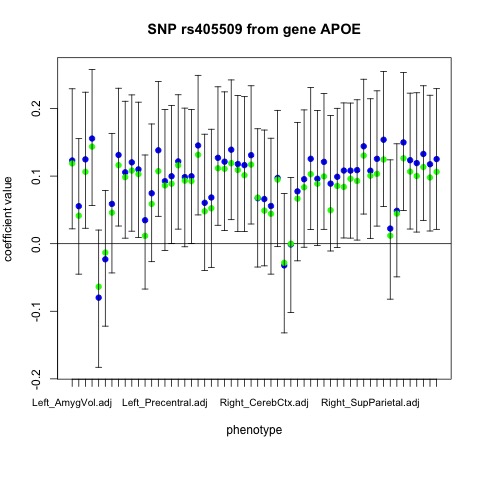}
\caption{Analysis of the ADNI data: The 95\% equal-tail credible intervals relating the SNP rs405509 from APOE to each of the $c=56$ imaging phenotypes along with the posterior mean estimate (blue) and the Wang et al. (2012) estimate (green). The imaging measures are represented on the x-axis in the same order as they are listed in the rows of Table 2, first for the left hemisphere followed by the right hemisphere.}
\end{figure}

\begin{figure}[!t]
\centering
\includegraphics[scale=0.7]{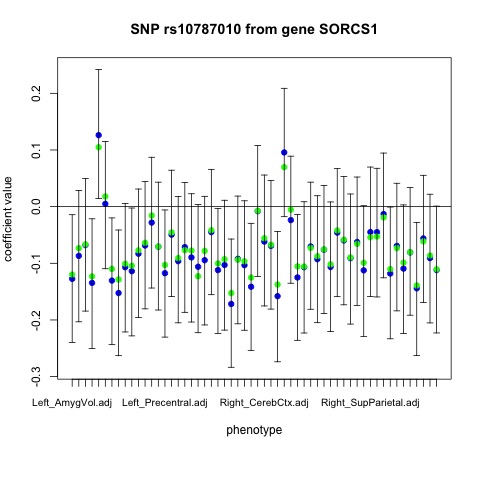}
\caption{Analysis of the ADNI data: The 95\% equal-tail credible intervals relating the SNP rs10787010 from the SORCS1 gene to each of the $c=56$ imaging phenotypes along with the posterior mean (blue) estimate and the Wang et al. (2012) estimate (green). The imaging measures are represented on the x-axis in the same order as they are listed in the rows of Table 2, first for the left hemisphere followed by the right hemisphere.}
\end{figure}

\tiny
\begin{longtable}{lllrrrr}
\caption{Analysis of ADNI Data - SNP-ROI regression roefficients with 95\% equal-tail credible interval excluding the value 0.}\\
SNP&Gene&ROI&Posterior.Mean&X95..CI.Lower&X95..CI.Upper&Wang et al. Estimate\tabularnewline
\hline
rs405509&APOE&Left AmygVol&$ 0.123$&$ 0.022$&$ 0.229$&$ 0.119$\tabularnewline
rs2327389&NEDD9&Left AmygVol&$ 0.189$&$ 0.004$&$ 0.383$&$ 0.080$\tabularnewline
rs10787010&SORCS1&Left AmygVol&$-0.127$&$-0.240$&$-0.014$&$-0.120$\tabularnewline
rs1473180&DAPK1&Left CerebCtx&$-0.198$&$-0.340$&$-0.055$&$-0.168$\tabularnewline
rs1433099&LDLR&Left CerebCtx&$ 0.212$&$ 0.058$&$ 0.365$&$ 0.182$\tabularnewline
rs12209631&NEDD9&Left CerebCtx&$ 0.165$&$ 0.020$&$ 0.318$&$ 0.092$\tabularnewline
rs1269918&SORCS1&Left CerebCtx&$ 0.155$&$ 0.026$&$ 0.286$&$ 0.118$\tabularnewline
rs405509&APOE&Left CerebWM&$ 0.125$&$ 0.023$&$ 0.224$&$ 0.106$\tabularnewline
rs11601726&GAB2&Left CerebWM&$ 0.145$&$ 0.004$&$ 0.287$&$ 0.077$\tabularnewline
rs2569537&LDLR&Left CerebWM&$-0.173$&$-0.316$&$-0.032$&$-0.103$\tabularnewline
rs1269918&SORCS1&Left CerebWM&$ 0.158$&$ 0.031$&$ 0.293$&$ 0.118$\tabularnewline
rs2418811&SORCS1&Left CerebWM&$ 0.201$&$ 0.020$&$ 0.380$&$ 0.127$\tabularnewline
rs4935774&SORL1&Left CerebWM&$-0.148$&$-0.265$&$-0.032$&$-0.114$\tabularnewline
rs405509&APOE&Left HippVol&$ 0.156$&$ 0.057$&$ 0.258$&$ 0.144$\tabularnewline
rs12209631&NEDD9&Left HippVol&$ 0.168$&$ 0.019$&$ 0.311$&$ 0.104$\tabularnewline
rs2756271&PRNP&Left HippVol&$ 0.121$&$ 0.013$&$ 0.230$&$ 0.102$\tabularnewline
rs10787010&SORCS1&Left HippVol&$-0.135$&$-0.250$&$-0.021$&$-0.123$\tabularnewline
rs10787010&SORCS1&Left InfLatVent&$ 0.126$&$ 0.014$&$ 0.242$&$ 0.105$\tabularnewline
rs1269918&SORCS1&Left InfLatVent&$-0.133$&$-0.264$&$-0.003$&$-0.085$\tabularnewline
rs2418811&SORCS1&Left InfLatVent&$-0.192$&$-0.375$&$-0.008$&$-0.116$\tabularnewline
rs11601726&GAB2&Left LatVent&$-0.155$&$-0.299$&$-0.021$&$-0.072$\tabularnewline
rs11191692&CALHM1&Left EntCtx&$-0.108$&$-0.218$&$-0.001$&$-0.071$\tabularnewline
rs1473180&DAPK1&Left EntCtx&$-0.191$&$-0.335$&$-0.049$&$-0.132$\tabularnewline
rs2756271&PRNP&Left EntCtx&$ 0.155$&$ 0.041$&$ 0.268$&$ 0.124$\tabularnewline
rs10787010&SORCS1&Left EntCtx&$-0.131$&$-0.244$&$-0.020$&$-0.110$\tabularnewline
rs10787011&SORCS1&Left EntCtx&$-0.148$&$-0.288$&$-0.008$&$-0.081$\tabularnewline
rs405509&APOE&Left Fusiform&$ 0.131$&$ 0.026$&$ 0.230$&$ 0.116$\tabularnewline
rs1473180&DAPK1&Left Fusiform&$-0.156$&$-0.300$&$-0.015$&$-0.107$\tabularnewline
rs10787010&SORCS1&Left Fusiform&$-0.152$&$-0.263$&$-0.041$&$-0.129$\tabularnewline
rs4311&ACE&Left InfParietal&$-0.212$&$-0.366$&$-0.056$&$-0.160$\tabularnewline
rs405509&APOE&Left InfParietal&$ 0.106$&$ 0.008$&$ 0.211$&$ 0.098$\tabularnewline
rs405509&APOE&Left InfTemporal&$ 0.120$&$ 0.018$&$ 0.220$&$ 0.108$\tabularnewline
rs2756271&PRNP&Left InfTemporal&$ 0.114$&$ 0.002$&$ 0.225$&$ 0.099$\tabularnewline
rs10787010&SORCS1&Left InfTemporal&$-0.114$&$-0.228$&$-0.002$&$-0.104$\tabularnewline
rs666004&SORL1&Left InfTemporal&$ 0.208$&$ 0.018$&$ 0.395$&$ 0.136$\tabularnewline
rs405509&APOE&Left MidTemporal&$ 0.110$&$ 0.009$&$ 0.209$&$ 0.103$\tabularnewline
rs6107516&PRNP&Left MidTemporal&$ 0.141$&$ 0.019$&$ 0.263$&$ 0.109$\tabularnewline
rs9314349&CLU&Left Parahipp&$-0.095$&$-0.187$&$-0.003$&$-0.049$\tabularnewline
rs6584307&ENTPD7&Left Parahipp&$ 0.148$&$ 0.010$&$ 0.289$&$ 0.088$\tabularnewline
rs1475345&NEDD9&Left Parahipp&$-0.185$&$-0.344$&$-0.033$&$-0.104$\tabularnewline
rs2756271&PRNP&Left Parahipp&$ 0.112$&$ 0.004$&$ 0.224$&$ 0.085$\tabularnewline
rs1473180&DAPK1&Left PostCing&$-0.169$&$-0.312$&$-0.029$&$-0.140$\tabularnewline
rs16924159&IL33&Left PostCing&$ 0.174$&$ 0.015$&$ 0.340$&$ 0.160$\tabularnewline
rs405509&APOE&Left Postcentral&$ 0.138$&$ 0.041$&$ 0.240$&$ 0.107$\tabularnewline
rs2149196&SORCS1&Left Postcentral&$-0.262$&$-0.477$&$-0.049$&$-0.174$\tabularnewline
rs1023024&SORCS1&Left Precentral&$ 0.182$&$ 0.009$&$ 0.366$&$ 0.088$\tabularnewline
rs10787010&SORCS1&Left Precentral&$-0.117$&$-0.230$&$-0.006$&$-0.103$\tabularnewline
rs1568400&THRA&Left Precentral&$ 0.111$&$ 0.009$&$ 0.214$&$ 0.092$\tabularnewline
rs7219773&TNK1&Left Precentral&$ 0.114$&$ 0.022$&$ 0.207$&$ 0.082$\tabularnewline
rs4311&ACE&Left Precuneus&$-0.237$&$-0.398$&$-0.079$&$-0.185$\tabularnewline
rs405509&APOE&Left Precuneus&$ 0.100$&$ 0.000$&$ 0.205$&$ 0.089$\tabularnewline
rs1433099&LDLR&Left Precuneus&$ 0.173$&$ 0.021$&$ 0.328$&$ 0.156$\tabularnewline
rs1699105&SORL1&Left Precuneus&$-0.156$&$-0.304$&$-0.003$&$-0.103$\tabularnewline
rs405509&APOE&Left SupFrontal&$ 0.122$&$ 0.022$&$ 0.220$&$ 0.116$\tabularnewline
rs744970&NEDD9&Left SupFrontal&$ 0.142$&$ 0.001$&$ 0.285$&$ 0.098$\tabularnewline
rs4311&ACE&Left SupParietal&$-0.235$&$-0.391$&$-0.076$&$-0.178$\tabularnewline
rs1556758&SORCS1&Left SupParietal&$-0.240$&$-0.465$&$-0.007$&$-0.156$\tabularnewline
rs4311&ACE&Left SupTemporal&$-0.210$&$-0.366$&$-0.057$&$-0.142$\tabularnewline
rs405509&APOE&Left SupTemporal&$ 0.100$&$ 0.000$&$ 0.199$&$ 0.093$\tabularnewline
rs405509&APOE&Left Supramarg&$ 0.145$&$ 0.043$&$ 0.249$&$ 0.132$\tabularnewline
rs17496723&NEDD9&Left Supramarg&$ 0.193$&$ 0.013$&$ 0.378$&$ 0.154$\tabularnewline
rs16924159&IL33&Left MeanCing&$ 0.184$&$ 0.024$&$ 0.346$&$ 0.149$\tabularnewline
rs10502262&SORL1&Left MeanCing&$ 0.205$&$ 0.014$&$ 0.395$&$ 0.130$\tabularnewline
rs1699105&SORL1&Left MeanCing&$-0.156$&$-0.304$&$-0.010$&$-0.095$\tabularnewline
rs405509&APOE&Left MeanFront&$ 0.127$&$ 0.027$&$ 0.232$&$ 0.112$\tabularnewline
rs744970&NEDD9&Left MeanFront&$ 0.154$&$ 0.012$&$ 0.296$&$ 0.102$\tabularnewline
rs10787010&SORCS1&Left MeanFront&$-0.112$&$-0.224$&$-0.003$&$-0.100$\tabularnewline
rs405509&APOE&Left MeanLatTemp&$ 0.121$&$ 0.019$&$ 0.225$&$ 0.111$\tabularnewline
rs405509&APOE&Left MeanMedTemp&$ 0.139$&$ 0.036$&$ 0.242$&$ 0.119$\tabularnewline
rs1473180&DAPK1&Left MeanMedTemp&$-0.193$&$-0.337$&$-0.049$&$-0.142$\tabularnewline
rs10787010&SORCS1&Left MeanMedTemp&$-0.172$&$-0.284$&$-0.057$&$-0.153$\tabularnewline
rs4311&ACE&Left MeanPar&$-0.220$&$-0.383$&$-0.064$&$-0.171$\tabularnewline
rs405509&APOE&Left MeanPar&$ 0.118$&$ 0.018$&$ 0.219$&$ 0.109$\tabularnewline
rs405509&APOE&Left MeanSensMotor&$ 0.116$&$ 0.018$&$ 0.218$&$ 0.101$\tabularnewline
rs1023024&SORCS1&Left MeanSensMotor&$ 0.183$&$ 0.007$&$ 0.359$&$ 0.087$\tabularnewline
rs2149196&SORCS1&Left MeanSensMotor&$-0.221$&$-0.445$&$-0.003$&$-0.125$\tabularnewline
rs7219773&TNK1&Left MeanSensMotor&$ 0.093$&$ 0.000$&$ 0.186$&$ 0.067$\tabularnewline
rs405509&APOE&Left MeanTemp&$ 0.131$&$ 0.029$&$ 0.234$&$ 0.117$\tabularnewline
rs1473180&DAPK1&Left MeanTemp&$-0.154$&$-0.299$&$-0.009$&$-0.114$\tabularnewline
rs10787010&SORCS1&Left MeanTemp&$-0.141$&$-0.254$&$-0.029$&$-0.125$\tabularnewline
rs4311&ACE&Right CerebCtx&$-0.172$&$-0.324$&$-0.016$&$-0.139$\tabularnewline
rs11141918&DAPK1&Right CerebCtx&$-0.194$&$-0.381$&$-0.009$&$-0.096$\tabularnewline
rs1473180&DAPK1&Right CerebCtx&$-0.160$&$-0.302$&$-0.015$&$-0.146$\tabularnewline
rs2025935&CR1&Right CerebWM&$-0.152$&$-0.290$&$-0.013$&$-0.139$\tabularnewline
rs16924159&IL33&Right CerebWM&$-0.164$&$-0.334$&$-0.002$&$-0.117$\tabularnewline
rs2569537&LDLR&Right CerebWM&$-0.152$&$-0.294$&$-0.011$&$-0.096$\tabularnewline
rs2418811&SORCS1&Right CerebWM&$ 0.198$&$ 0.020$&$ 0.379$&$ 0.128$\tabularnewline
rs4935774&SORL1&Right CerebWM&$-0.180$&$-0.296$&$-0.064$&$-0.134$\tabularnewline
rs12209631&NEDD9&Right HippVol&$ 0.157$&$ 0.013$&$ 0.301$&$ 0.086$\tabularnewline
rs7938033&PICALM&Right HippVol&$-0.167$&$-0.330$&$-0.002$&$-0.125$\tabularnewline
rs2756271&PRNP&Right HippVol&$ 0.126$&$ 0.017$&$ 0.242$&$ 0.111$\tabularnewline
rs10787010&SORCS1&Right HippVol&$-0.158$&$-0.274$&$-0.044$&$-0.138$\tabularnewline
rs10787011&SORCS1&Right HippVol&$-0.138$&$-0.273$&$-0.001$&$-0.111$\tabularnewline
rs2025935&CR1&Right InfLatVent&$ 0.199$&$ 0.060$&$ 0.341$&$ 0.155$\tabularnewline
rs3095747&DAPK1&Right InfLatVent&$-0.162$&$-0.289$&$-0.038$&$-0.126$\tabularnewline
rs928413&IL33&Right InfLatVent&$-0.134$&$-0.268$&$-0.001$&$-0.055$\tabularnewline
rs4305&ACE&Right LatVent&$ 0.143$&$ 0.009$&$ 0.280$&$ 0.076$\tabularnewline
rs7938033&PICALM&Right EntCtx&$-0.185$&$-0.358$&$-0.019$&$-0.113$\tabularnewline
rs10787010&SORCS1&Right EntCtx&$-0.125$&$-0.236$&$-0.014$&$-0.106$\tabularnewline
rs10787011&SORCS1&Right EntCtx&$-0.141$&$-0.282$&$-0.003$&$-0.106$\tabularnewline
rs2025935&CR1&Right Fusiform&$-0.144$&$-0.287$&$-0.005$&$-0.121$\tabularnewline
rs4311&ACE&Right InfParietal&$-0.197$&$-0.356$&$-0.031$&$-0.151$\tabularnewline
rs405509&APOE&Right InfParietal&$ 0.126$&$ 0.021$&$ 0.231$&$ 0.103$\tabularnewline
rs3118846&DAPK1&Right InfParietal&$ 0.173$&$ 0.001$&$ 0.349$&$ 0.179$\tabularnewline
rs10502262&SORL1&Right InfTemporal&$ 0.193$&$ 0.005$&$ 0.378$&$ 0.113$\tabularnewline
rs405509&APOE&Right MidTemporal&$ 0.121$&$ 0.021$&$ 0.222$&$ 0.099$\tabularnewline
rs6107516&PRNP&Right MidTemporal&$ 0.121$&$ 0.000$&$ 0.244$&$ 0.088$\tabularnewline
rs4311&ACE&Right PostCing&$-0.172$&$-0.332$&$-0.015$&$-0.140$\tabularnewline
rs17399090&DAPK1&Right PostCing&$ 0.189$&$ 0.043$&$ 0.338$&$ 0.135$\tabularnewline
rs3124237&DAPK1&Right PostCing&$-0.165$&$-0.332$&$-0.007$&$-0.141$\tabularnewline
rs212539&ECE1&Right PostCing&$-0.214$&$-0.424$&$-0.009$&$-0.098$\tabularnewline
rs12248379&SORCS1&Right PostCing&$ 0.188$&$ 0.027$&$ 0.362$&$ 0.092$\tabularnewline
rs4311&ACE&Right Postcentral&$-0.238$&$-0.396$&$-0.080$&$-0.172$\tabularnewline
rs405509&APOE&Right Postcentral&$ 0.108$&$ 0.008$&$ 0.208$&$ 0.084$\tabularnewline
rs4878117&DAPK1&Right Postcentral&$-0.134$&$-0.259$&$-0.008$&$-0.080$\tabularnewline
rs2149196&SORCS1&Right Postcentral&$-0.231$&$-0.448$&$-0.014$&$-0.125$\tabularnewline
rs3744805&THRA&Right Postcentral&$ 0.177$&$ 0.051$&$ 0.306$&$ 0.112$\tabularnewline
rs7219773&TNK1&Right Postcentral&$ 0.111$&$ 0.018$&$ 0.204$&$ 0.075$\tabularnewline
rs4311&ACE&Right Precentral&$-0.175$&$-0.330$&$-0.014$&$-0.139$\tabularnewline
rs405509&APOE&Right Precentral&$ 0.108$&$ 0.008$&$ 0.208$&$ 0.096$\tabularnewline
rs3744805&THRA&Right Precentral&$ 0.166$&$ 0.040$&$ 0.295$&$ 0.112$\tabularnewline
rs4311&ACE&Right Precuneus&$-0.230$&$-0.386$&$-0.073$&$-0.180$\tabularnewline
rs405509&APOE&Right Precuneus&$ 0.109$&$ 0.005$&$ 0.213$&$ 0.093$\tabularnewline
rs3811450&CHRNB2&Right Precuneus&$-0.133$&$-0.269$&$-0.001$&$-0.115$\tabularnewline
rs3124237&DAPK1&Right Precuneus&$-0.182$&$-0.342$&$-0.015$&$-0.110$\tabularnewline
rs1433099&LDLR&Right Precuneus&$ 0.159$&$ 0.003$&$ 0.318$&$ 0.139$\tabularnewline
rs4311&ACE&Right SupFrontal&$-0.237$&$-0.392$&$-0.081$&$-0.181$\tabularnewline
rs405509&APOE&Right SupFrontal&$ 0.144$&$ 0.044$&$ 0.243$&$ 0.130$\tabularnewline
rs3124237&DAPK1&Right SupFrontal&$-0.177$&$-0.343$&$-0.014$&$-0.112$\tabularnewline
rs4311&ACE&Right SupParietal&$-0.254$&$-0.410$&$-0.099$&$-0.191$\tabularnewline
rs405509&APOE&Right SupParietal&$ 0.108$&$ 0.008$&$ 0.214$&$ 0.101$\tabularnewline
rs405509&APOE&Right SupTemporal&$ 0.126$&$ 0.026$&$ 0.226$&$ 0.103$\tabularnewline
rs405509&APOE&Right Supramarg&$ 0.154$&$ 0.055$&$ 0.255$&$ 0.125$\tabularnewline
rs10502262&SORL1&Right Supramarg&$ 0.193$&$ 0.004$&$ 0.381$&$ 0.117$\tabularnewline
rs10787010&SORCS1&Right TemporalPole&$-0.118$&$-0.233$&$-0.001$&$-0.110$\tabularnewline
rs1568400&THRA&Right TemporalPole&$-0.106$&$-0.212$&$-0.004$&$-0.088$\tabularnewline
rs17399090&DAPK1&Right MeanCing&$ 0.187$&$ 0.038$&$ 0.336$&$ 0.142$\tabularnewline
rs4311&ACE&Right MeanFront&$-0.200$&$-0.361$&$-0.044$&$-0.156$\tabularnewline
rs405509&APOE&Right MeanFront&$ 0.150$&$ 0.049$&$ 0.253$&$ 0.126$\tabularnewline
rs405509&APOE&Right MeanLatTemp&$ 0.124$&$ 0.022$&$ 0.223$&$ 0.107$\tabularnewline
rs405509&APOE&Right MeanMedTemp&$ 0.119$&$ 0.017$&$ 0.223$&$ 0.100$\tabularnewline
rs10787010&SORCS1&Right MeanMedTemp&$-0.144$&$-0.263$&$-0.028$&$-0.139$\tabularnewline
rs1699102&SORL1&Right MeanMedTemp&$-0.246$&$-0.481$&$-0.020$&$-0.110$\tabularnewline
rs4311&ACE&Right MeanPar&$-0.240$&$-0.395$&$-0.082$&$-0.183$\tabularnewline
rs405509&APOE&Right MeanPar&$ 0.133$&$ 0.035$&$ 0.234$&$ 0.113$\tabularnewline
rs4311&ACE&Right MeanSensMotor&$-0.210$&$-0.367$&$-0.052$&$-0.161$\tabularnewline
rs405509&APOE&Right MeanSensMotor&$ 0.118$&$ 0.019$&$ 0.220$&$ 0.098$\tabularnewline
rs4878117&DAPK1&Right MeanSensMotor&$-0.133$&$-0.261$&$-0.007$&$-0.080$\tabularnewline
rs3744805&THRA&Right MeanSensMotor&$ 0.185$&$ 0.058$&$ 0.313$&$ 0.123$\tabularnewline
rs4311&ACE&Right MeanTemp&$-0.160$&$-0.316$&$-0.003$&$-0.126$\tabularnewline
rs405509&APOE&Right MeanTemp&$ 0.125$&$ 0.021$&$ 0.229$&$ 0.106$\tabularnewline
rs1699102&SORL1&Right MeanTemp&$-0.245$&$-0.468$&$-0.024$&$-0.117$\tabularnewline
\hline
\end{longtable}

\end{document}